\documentclass{aa}
\usepackage{graphicx}
\usepackage{txfonts}
\usepackage{natbib}
\bibpunct{(}{)}{;}{a}{}{,}
\begin{document}
   \title{Direct evidence of a sub-stellar companion around CT Cha\thanks{Based on observations made with ESO telescopes at the Paranal Observatory under program IDs 076.C-0292(A), 078.C-0535(A), \& 279.C-5010(A)}\fnmsep \thanks{Color versions of Figs.~4, 6, 8, 11 are only available in electronic form at http://www.aanda.org}}
   \titlerunning{Evidence of a co-moving sub-stellar companion of CT Cha}

   \author{T. O.\,B. Schmidt
          \inst{1}
          \and
          R. Neuh\"auser\inst{1}
          \and
          A. Seifahrt\inst{1,2}
          \and
          N. Vogt\inst{3,4}
          \and
          A. Bedalov\inst{1,5}
          \and
          Ch. Helling\inst{6}
          \and
          S. Witte\inst{7}
          \and
          P. H. Hauschildt\inst{7}
          }

   \offprints{Tobias Schmidt, e-mail:~tobi@astro.uni-jena.de}

   \institute{Astrophysikalisches Institut und Universit\"ats-Sternwarte, Universit\"at Jena, Schillerg\"a\ss chen 2-3, 07745 Jena, Germany\\
              \email{tobi@astro.uni-jena.de}
         \and
             Institut f\"ur Astrophysik, Universit\"at G\"ottingen, Friedrich-Hund-Platz 1, 37077 G\"ottingen, Germany
         \and
             Departamento de F\'isica y Astronom\'ia, Universidad de Valpara\'iso, Avenida Gran Breta\~na 1111, Valpara\'iso, Chile
	 \and
	     Instituto de Astronom\'ia, Universidad Catolica del Norte, Avda.~Angamos 0610, Antofagasta, Chile
         \and
             Faculty of Natural Sciences, University of Split, Teslina 12. 21000 Split, Croatia
         \and
             SUPA, School of Physics and Astronomy, University of St. Andrews, North Haugh, St. Andrews KY16 9SS, UK
         \and
             Hamburger Sternwarte, Gojenbergsweg 112, 21029 Hamburg, Germany
             }

   \date{Received October 2007; accepted August 2008}

 
  \abstract
   {}
   {In our ongoing search for close and faint companions around T Tauri stars in the Chamaeleon star-forming region, we here present observations of a new common proper motion companion to the young T-Tauri star and Chamaeleon member CT Cha and discuss its properties in comparison to other young, low-mass objects and to synthetic model spectra from different origins.}
   {Common proper motion of the companion and CT Cha was confirmed by direct Ks-band imaging data taken with the VLT Adaptive Optics (AO) instrument NACO in February 2006
    and March 2007, together with a Hipparcos binary for astrometric calibration. An additional J-band image was taken in March 2007 to obtain color information for a first classification of the companion. Moreover, AO integral field spectroscopy with SINFONI in J, and H+K bands was obtained to deduce physical parameters of the companion, such as temperature and extinction. Relative flux calibration of the bands was achieved using photometry from the NACO imaging data.}
   {We found a very faint (Ks\,=\,14.9 mag, Ks$_{0}$\,=\,14.4 mag) object, just $\sim$\,2.67\arcsec~northwest of CT Cha corresponding to a projected separation of $\sim$\,440\,AU at 165\,$\pm$\,30\,pc. We show that CT Cha A and this faint object form a common proper motion pair and that the companion is by $\geq$\,4\,$\sigma$ significance not a stationary background object. The near-infrared spectroscopy yields a temperature of 2600\,$\pm$\,250\,K for the companion and an optical extinction of $A_{\rm V}$=\,5.2\,$\pm$\,0.8\,mag, when compared to spectra calculated from Drift-Phoenix model atmospheres. We demonstrate the validity of the model fits by comparison to several other well-known young sub-stellar objects.}
   {We conclude that the CT Cha companion is a very low-mass member of Chamaeleon and very likely a physical companion to CT Cha, as the probability for a by chance alignment is $\leq$\,0.01. Due to a prominent Pa-$\beta$ emission in the J-band, accretion is probably still ongoing onto the CT Cha companion. From temperature and luminosity ($\log(L_{bol}/L_{\odot})$=\,--2.68\,$\pm$\,0.21), we derive a radius of R=\,2.20$_{-0.60}^{+0.81}$\,R$_{\mathrm{Jup}}$. We find a consistent mass of M=\,17\,$\pm$\,6\,M$_{\mathrm{Jup}}$ for the CT Cha companion from both its luminosity and temperature when placed on evolutionary tracks. Hence, the CT Cha companion is most likely a wide brown dwarf companion or possibly even a planetary mass object.}


   \maketitle
%

\section{Introduction} 


\object{CT Cha} (aka HM\,9), introduced in the 65th Name-List of Variable stars by 
\citet{1981IBVS.1921....1K}, was originally found by \citet{1973ApJ...180..115H} as 
an emission-line star in Chamaeleon, exhibiting variations in its H$\alpha$\,line 
from plate to plate and showing partial veiling \citep{1980AJ.....85..444R}.

While the star was first classified as a T Tauri star by \citet{1987MNRAS.224..497W},
it was later found to be a classical T Tauri star by \citet{1990ApJS...74..575W} 
and \citet{1992ApJ...385..217G} from IRAS data. \citet{2000ApJ...534..838N} found 
evidence of a silicate feature disk ($L_{sil}=10^{-2}\,L_{\odot}$, $L_{sil}/L_{\ast}=0.014$), 
using ISO data.

The variations in the H$\alpha$\,line were later interpreted as accretion signatures 
when \citet{1998ApJ...495..385H} measured a mass accretion rate of log{\,\.{M}}$=-8.28\,M_{\odot}/a$.
Additional variations in infrared \citep{1979MNRAS.187..305G} and optical photometry can 
possibly be explained by surface features on CT Cha at a rotation period of 9.86 days,
as found by \citet{1998A&AS..128..561B}.

All additional properties of the K7 \citep{1988A&AS...76..347G} star CT Cha, such as
its age ranging from 0.9 Myr \citep{2000ApJ...534..838N} to 3 Myr \citep{1993ApJ...416..623F},
as well as its equivalent width of the lithium absorption line of W$_{\lambda}$(Li)\,=\,0.40\,$\pm$\,0.05\,\AA\,
\citep{2007A&A...467.1147G}, its radial velocity of 15.1\,$\pm$\,0.1 km/s \citep{2006A&A...448..655J} 
and proper motion (Table\,\ref{table:2}), are consistent with a very young member of the Cha I star-forming 
region, having an age of 2\,$\pm$\,2\,Myr.

\section{Direct observations of a wide companion}

\subsection{AO imaging detection}

We observed CT Cha in two epochs in February 2006 and in March 2007 (see Table \ref{table:1}).
All observations were done with the European Southern Observatory (ESO) Very Large Telescope (VLT) instrument Naos-Conica
\citep[NACO,][]{2003SPIE.4841..944L,2003SPIE.4839..140R}.
In all cases the S13 camera ($\sim$13 mas/pixel) was used
in double-correlated read-out mode.

\begin{figure}
   \resizebox{\hsize}{!}{\includegraphics{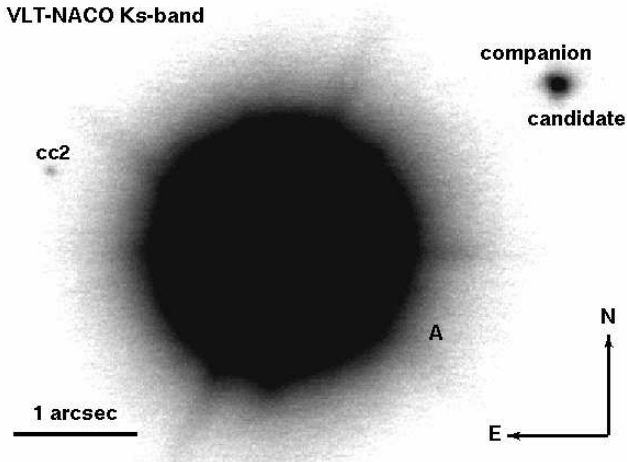}}
   \caption{VLT-NACO Ks-band image of CT Cha and its 6.3 mag fainter companion candidate (2.670\,$\pm$\,0.038\arcsec northwest) 
   from 2007 March 2. The object marked as `cc2' was found to be a background object.}
   \label{image}
   \end{figure}

For the raw data reduction, we subtracted a mean dark from all science frames and the flatfield frames,
then divided by the normalized dark-subtracted flatfield, and subtracted the mean background by using
ESO \textit{eclipse\,/\,jitter}.
In all three images a companion candidate 
was found 2.67\arcsec~northwest of CT Cha (Fig.~\ref{image}), 
corresponding to $\sim$\,440 AU at a distance of 165\,$\pm$\,30 pc, 
the latter estimated from the combination of data by \citet{1999A&A...352..574B} and \cite{1997A&A...327.1194W} 
for Cha I members.

\begin{table}
\caption{VLT/NACO observation log}
\label{table:1}
\begin{tabular}{lcccccccc}
\hline\hline
JD - 2453700          & Date of     & DIT & NDIT & No.~of & Filter\\
$[\mathrm{days}]$     & observation & [s] &      & images  &       \\
\hline
\ \ 83.54702 & 17 Feb 2006 & 1.5 & 25 & 20 & Ks \\
   460.62535 &  1 Mar 2007 & 4   & 15 & 21 & J  \\
   461.63001 &  2 Mar 2007 & 4   & 7  & 30 & Ks \\
\hline
\end{tabular}
Remark: Each image consists of the number of exposures given in column 4 times the individual integration time given in column 3.
\end{table}

\subsection{Astrometry}

   \begin{figure}
   \resizebox{\hsize}{!}{\includegraphics{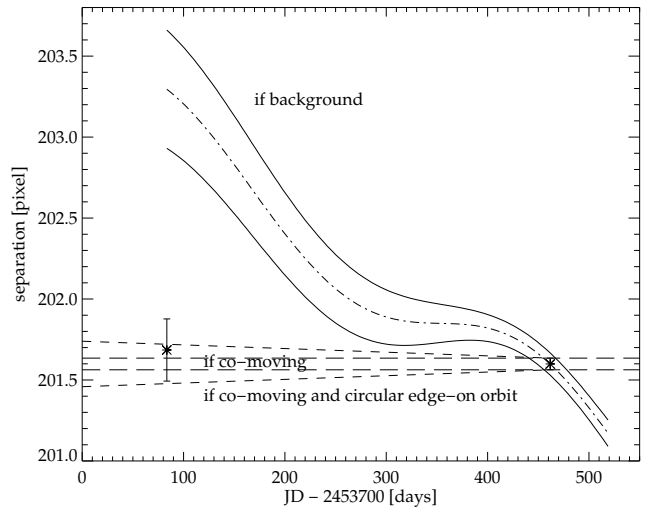}}
   \caption{Observed separation between CT Cha A and the CT Cha companion. Our two measurements from 2006 and 2007 are shown. 
            The long dashed lines enclose the area for constant separation, as expected for a co-moving object. 
            The dash-dotted line is the change expected if the CT Cha companion is a non-moving background star. The opening cone 
            enclosed by the continuous lines its estimated errors. The waves of this cone show the differential 
            parallactic motion that has to be taken into account if the other component is a non-moving background star 
            with negligible parallax. The opening short-dashed cone is for the combination of co-motion and the maximum 
            possible orbital motion for a circular edge-on orbit.}
   \label{Sep}
   \end{figure}

To check for the common proper motion of the tentative companion of CT Cha, 
we used the proper motion 
(PM) of the star published in the literature (Table ~\ref{table:2}). 
We used the weighted mean proper motion for checking whether the two objects show common PM below.
To determine the positions of both components, 
we constructed a reference PSF from both objects. Thus, we obtained an appropriate reference PSF for each single image. 
Using IDL/starfinder, we scaled and shifted the reference PSF simultaneously to both components in each of our individual 
images by minimizing the residuals.
Realistic error estimates in position and flux for each object were obtained from the mean and standard deviation of the positions found in all individual images of
a single epoch.

\begin{table}
\caption{Proper motions of CT Cha}
\label{table:2}
\begin{center}
\begin{tabular}{lr@{\,$\pm$\,}lr@{\,$\pm$\,}l}
\hline\hline
Reference &\multicolumn{2}{c}{$\mu_{\alpha} \cos{\delta}$}    & \multicolumn{2}{c}{$\mu_{\delta}$}  \\
          &\multicolumn{2}{c}{[mas/yr]} & \multicolumn{2}{c}{[mas/yr]}\\
\hline
UCAC2 \citep{2004AJ....127.3043Z}          & -22.2  & 5.2  & 7    & 5.2 \\
ICRF ext. \citep{Camargo2003}              & -18    & 10   & 4    & 9   \\
\hline
weighted mean                              & -21.3  & 4.6  & 6.3  & 4.5  \\
\hline
\end{tabular}
\end{center}
\end{table}

We calibrated the NACO data using the wide binary star HIP 73357 for our two measurements in 2006 and 2007.
The astrometry of this binary was measured very accurately by the Hipparcos satellite. However, 15~years have passed
since these measurements, allowing a large possible orbital motion of the binary, which is now dominating the astrometric uncertainty,
resulting in the calibration given in Table~\ref{table:3}. The astrometric errors include Hipparcos errors, maximum possible orbital motion of the calibration binary and measurement errors of the position of CT Cha and its companion. Our derived pixel scale is in good agreement with earlier measurements
as e.g. in \citet{2005A&A...435L..13N, 2008A&A...484..281N}. 

The common proper motion analysis made use of relative measurements; hence, the astrometric calibration used for our images took into account only the uncertainties in separation and position angle, as well as in proper motion and orbital motion between the two epochs of 2006 and 2007. This allows precise measurement of the relative motions of CT Cha and the companion candidate, given in Table~\ref{table:4}. The absolute astrometry, which has to take the full uncertainties of the astrometric calibration into account, is given in Table~\ref{table:3}.
We would like to stress that common proper motion can be shown based on relative astrometry, while absolute values are given for future comparisons, also with different instruments.

\begin{table}
\caption{Absolute astrometric results for CT Cha A and companion}
\label{table:3}
\centering
\begin{tabular}{lccr@{\,$\pm$\,}l}
\hline\hline
JD - 2453700 & Pixel scale & Separation  & \multicolumn{2}{c}{PA$^a$ }\\
$[\mathrm{days}]$ & [mas/Pixel] & [arcsec] &  \multicolumn{2}{c}{[$\deg$]} \\
\hline
 \ \ 83.54702 & 13.24 $\pm$ 0.18 & 2.670 $\pm$ 0.036 & 300.71 & 1.24\\
   461.63001  & 13.24 $\pm$ 0.19 & 2.670 $\pm$ 0.038 & 300.68 & 1.32\\
\hline
\end{tabular}
\begin{flushleft}
Remarks: All data from Ks-band images. (a) PA is measured from N over E to S.
\end{flushleft}
\end{table}

From Fig.~\ref{Sep}, we can exclude by 3.9\,$\sigma$ that the CT Cha companion is a non-moving background object. Due to the location of the companion (northwest of CT Cha A) and the proper motion (also towards northwest), the position angle (PA)
gives only little additional significance (1.2\,$\sigma$ deviation from the background hypothesis), resulting in a combined significance of $\geq$\,4\,$\sigma$ for the exclusion of the background hypothesis. For a K7 star and a sub-stellar companion (see below) at a projected separation of $\sim$\,440 AU (at $\sim165$ pc), the orbital period is $\sim 11000$ yrs;
hence, the maximum change in separation due to orbital motion (for circular edge-on orbit) is $\sim 1$ mas/yr ($\sim 0.07$ pix/yr) or $\sim 0.03\,^{\circ}$/yr in PA for pole-on orbit.
Neither in separation nor in position angle 
any significant sign of orbital motion could be detected, 
given the short epoch difference ($\sim$\,1 year).

\begin{table}
\caption{Relative astrometric results for CT Cha A and companion}
\label{table:4}
\centering
\begin{tabular}{ccc}
\hline\hline
Epoch differ- & Change in sepa-& Change in \\
ence [days] &  ration [pixel] & PA$^a$ [$\deg$] \\
\hline
 378.08299        & -0.09 $\pm$ 0.20 & -0.036  $\pm$ 0.086\\
\hline
\hline
\end{tabular}
\begin{flushleft}
Remarks: All data from Ks-band images. (a) PA is measured from N over E to S.
\end{flushleft}
\end{table}

\begin{table}
\caption{Apparent magnitudes of the CT Cha companion}
\label{table:5}
\centering
\begin{tabular}{lr@{\,$\pm$\,}lcc}
\hline\hline
Epoch & \multicolumn{2}{c}{J-band} & Ks-band \\
\hline
17 Feb 2006   & --     & --             & 14.95 $\pm$ 0.30\\
1/2 Mar 2007  & 16.61 & 0.30 & 14.89 $\pm$ 0.30\\
\hline
\end{tabular}
\end{table}

While the negligible differences seen in PA and separation between the different observations are consistent with common proper motion, a possible difference in proper motion between both objects of up to a few mas/yr cannot be excluded from the data. 
Such a difference in proper motion 
would be typical of the velocity dispersion in star-forming regions like Cha I \citep{2005A&A...438..769D}.
As a result we cannot yet exclude both objects being independent members of Cha I, thus, not orbiting each other.
Even if this were the case, the age and distance (within the given uncertainties) would be the same for both objects,
and likewise the mass estimation. However, the probability of finding such a red object at a projected separation less 
than 2.67\arcsec~but physically unrelated to CT Cha is very small. Taking the number density of objects with similar
spectral type and same (or higher) brightness from the \textit{Dwarf Archives}\footnote{www.DwarfArchives.org}, we find a probability
of 4 $\cdot 10^{-9}$ that an object like the CT Cha companion within 2.67\arcsec~is a chance alignment. Even in the denser areas of the Chamaeleon I star-forming region, objects that red are scarce, and the number density obtained from \citet{2006ApJ...649..894L} results in a probability of approximately 4\,$\cdot 10^{-3}$ for a chance alignment.
Following the approach of \citet{2008arXiv0803.0561L}, we can estimate the probability of a chance alignment from the surface density of all sources surrounding CT Cha from the Two Micron All Sky Survey (2MASS) Point Source Catalogue (PSC) of the same brightness (or brighter) than our companion candidate within the same angular separation. This calculation results in a probability of 8\,$\cdot 10^{-3}$, while one should be careful since the completeness of 2MASS is only assured until $\sim$ 15\,mag in Ks-band, which is comparable to the apparent magnitude of the CT Cha companion. It is thus likely that the two objects are orbiting each other. Even if not bound, they are both young, hence members of Chamaeleon, i.e.~at roughly the same distance and age.

The presence of an additional source that can be ruled out as companion due to its color and proper motion (see `cc2' in Fig.~\ref{image}), can be explained by the additional source being approximately 2 magnitudes fainter than our newly found co-moving companion and the number of sources in the 2MASS PSC being increasing logarithmically with their brightness until the completeness limit of the catalog, resulting in a drastically increased surface density of such fainter objects if the number of sources is extrapolated to lower brightnesses.

\subsection{Photometry}

As described in the last section, we also obtained from the PSF fitting of both components the flux ratio of the CT Cha companion and CT Cha A.
Using the photometry of CT Cha A from the Two Micron All Sky Survey (2MASS) catalog of $J$\,=\,9.715\,$\pm$\,0.024\,mag, and $K$\,=\,8.661\,$\pm$\,0.021\,mag, and adding 0.3\,mag variability in A estimated from data by \citet{1998A&AS..128..561B}, \cite{1997ApJ...481..378G}, and \cite{1996MNRAS.280.1071L} to the error of the companion, we obtain its photometry (\ref{table:5}).

The Ks-band magnitudes obtained in the two epochs agree within their 1$\sigma$ errors, giving no indication of photometric variability. From the extinction-corrected $J_{0}\,-\,Ks_{0}$=\,0.84\,$\pm\,$0.50\,mag (2007, see Sect. 4) we estimate a spectral type of earlier than L5 for the faint CT Cha companion using the dwarf scale in \citet{2004AJ....127.3516G}.

\section{Spectroscopy}

In addition to our NACO measurements, we used the adaptive-optics integral-field spectrograph SINFONI, also mounted at UT 4 of the ESO VLT, to obtain spectra that do not suffer from wavelength-dependent slit losses that occur on normal spectrographs with narrow entrance slits, especially when combined with AO \citep{2003SPIE.4839.1117G}.

The observations of the CT Cha companion were carried out in H+K (resolution 1500) and J (resolution 2000) band in the nights of 16 \& 17 May 2007. Six nodding cycles with an integration time of 277s per frame were obtained each. We chose the maximum possible pixel scale (50 x 100 mas) of the instrument leaving CT Cha A as AO guide star outside of the FoV (3\arcsec~$\times$ 3\arcsec). All observations were done at the best possible airmass for the target ($\sim$\,1.6) and good seeing conditions of $\sim$\,0.6\arcsec~(optical DIMM seeing).

As telluric standards HIP 54257, a B6 IV/V star, and HIP 75445, a G1 -- G3 V star, were used for the J-band and H+K-band, respectively. In order to correct for features of these standard stars, the Pa-$\beta$ absorption at $\sim$\,1.282 $\mu$m of HIP 54257 was fitted by a Lorentzian profile and removed by division, and the metal lines and molecular bands of the sun-like star HIP 75445 were removed by a convolved high-resolution solar spectrum made available by the NSO/Kitt Peak Observatory.

We used the SINFONI data reduction pipeline version 1.7.1 offered by ESO \citep{2006ASPC..351..295J} with reduction routines developed by the SINFONI consortium \citep{2006NewAR..50..398A}. After standard reduction, all nodding cycles were combined to a final data cube. We used the \textit{Starfinder} package of IDL \citep{2000SPIE.4007..879D} and an iterative algorithm to remove the halo of CT Cha A, both described in detail in \citet{2007A&A...463..309S}. Since the J-band and H+K-band spectra do not overlap, we convolved the NACO and 2MASS filter transmission curves in a final reduction step with our measured spectra in J- and Ks-band and aligned the relative flux levels of both spectra by comparison of the integrated flux with the flux in J- and Ks-band of the photometry from March 1\,/\,2 2007 (see Table \ref{table:5}) converted to a flux density using the \textit{Spitzer Science Center Magnitude to Flux Density converter}\footnote{http://ssc.spitzer.caltech.edu/tools/magtojy/} based on \citet{2003AJ....126.1090C} for the calibration of 2MASS.

To put the CT Cha companion into context, we compare its spectra to model atmospheres, as well as to previously found sub-stellar companions and free-floating objects in very young associations in the following sections.

\subsection{Comparing with Tsuji {\sc Unified Cloudy} and with {\sc GAIA Dusty-Phoenix} model atmospheres}

We compare our spectrum with synthetic model spectra, which have to account for dust, beginning to condensate at temperatures T$_{\rm eff}$\,$\approx$\,2700~K in the outer atmospheric layers \citep{1996A&A...308L..29T,2005MNRAS.361.1323C}.
Therefore we used the GAIA {\sc Dusty-Phoenix} models v2.0 \citep{2005ESASP.576..565B} and Tsuji's unified cloudy models (UCM) with different values of T$_{\mathrm{\rm crit}}$ \citep{2005ApJ...621.1033T}. Tsuji's UCM-models were originally intended for old L- and T-dwarfs, where T$_{\mathrm{\rm crit}}$ was introduced to parametrize a possibly varying cloud height. The Tsuji UCM-models were calculated for lower surface gravities for this work by Tsuji (private communication).

Because none of these models fitted our data across the whole parameter range, we tried to account for an additional broad-band opacity contribution in the frame of these models. As shown in Fig.~6 of \citet{2006ApJ...640.1063B}, dwarf spectra strongly depend on the cloud particle size and the actual dust chemistry, but the grain sizes change depending on altitude \citep{2006A&A...455..325H}. Due to the unavailability of these models for low surface gravities, we simulated the broad veiling in the near-IR spectra and the partial filling of the spectral troughs and flattening of the otherwise strongly peaked H-band due to the presence of more clouds of small grains by adding an additional blackbody of similar effective temperature as compared to the underlying synthetic models.

The best fit could be achieved using the modified UCM grid for T$_{\mathrm{\rm eff}}$=\,2200\,$\pm$\,200\,K (T$_{\mathrm{crit}}$=\,1800\,K), 
a visual extinction of A$_{V}$=\,2.8\,$\pm$\,0.8\,mag in combination with an additional blackbody of 2250\,$\pm$\,250\,K, at a flux ratio of 1:1 with respect to the unmodified UCM spectra. This corresponds to a spectral type of M9 -- L3 using the dwarf spectral type to temperature conversion in \citet{2004AJ....127.3516G}.
However, all derived parameters are highly correlated, and we cannot prove that an additional blackbody accounts for an additional broad-band opacity source.

\subsection{Comparing with {\sc Drift-Phoenix} model atmospheres}

In contrast to previous model atmosphere results like {\sc Dusty-} and {\sc Cond-Phoenix}, the Tsuji-UCMs, the {\sc Settle-Phoenix} \citep{2007A&A...474L..21A} and the Ackerman \& Marley models \citep{2001ApJ...556..872A, 2007ApJ...655..541M}, \citet{2003A&A...399..297W, 2004A&A...414..335W}, and \citet{2006A&A...455..325H} presented a kinetic model description of the formation and evolution of dust in brown dwarfs. These models are studied in a test case comparison in \citet{2008IAUS..249..173H}.

Recently, \citet{2007IAUS..239..227D} and \citet{2008ApJ...675L.105H} combined the non-equilibrium, stationary cloud model from \citet[][{\sc DRIFT}: Nucleation, seed formation, growth, evaporation, gravitational settling, convective overshooting\,/\,up-mixing, element conservation]{2008arXiv0803.4315H} with a general-purpose model atmosphere code \citep[][{\sc PHOENIX}: Radiative transfer, hydrostatic equilibrium, mixing length theory, chemical equilibrium]{1999JCoAM.109...41H}. This gives us the unique opportunity to compare and classify spectra of low-temperature objects, where the formation of dust-cloud layers play a major role in the spectral appearance.

We used an almost complete grid of {\sc Drift-Phoenix} models in the range of T$_{\rm eff}$=\,2000\,$\ldots$\,2800\,K, $\log{g}$=\,3.0\,$\ldots$\,4.0, and [M/H]=\,-0.5\,$\ldots$\,0.5 in steps of 100\,K, 0.5, and 0.5, respectively. Moreover, we still need to account for reddening of our spectra, which increases the importance of the relative calibration of our spectra in J-band and H+K-band, as the accuracy of the extinction determination increases with the spectral coverage. This correction is important as the effective temperature is highly correlated with the extinction, because both values change the slope of the spectrum in J-, H-, and K-band, and already less than $A_{V}$\,=\,2\,mag can change the shape of the spectra significantly \citep{2008MNRAS.383.1385L}.

   \begin{figure}
   \resizebox{\hsize}{!}{\includegraphics{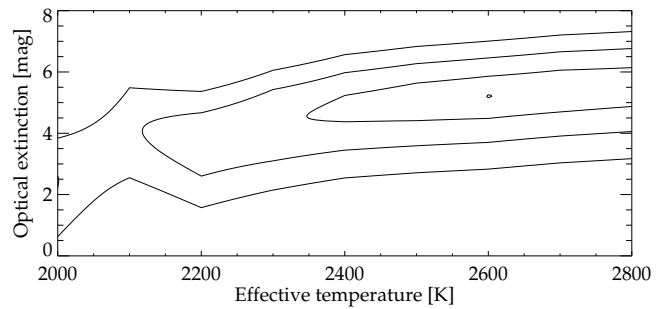}}
   \caption{Result of the $\chi^2$ minimization analysis for the CT Cha companion. Plotted are the best value (\textit{point}) and the 1, 2, and 3 sigma error contours for effective temperature T$_{\rm eff}$ and optical extinction $A_{V}$, determined from comparison of our SINFONI spectrum and the Drift-Phoenix model grid.}
   \label{chi_CTCha}
   \end{figure}

   \begin{figure*}
   \resizebox{\hsize}{!}{\includegraphics{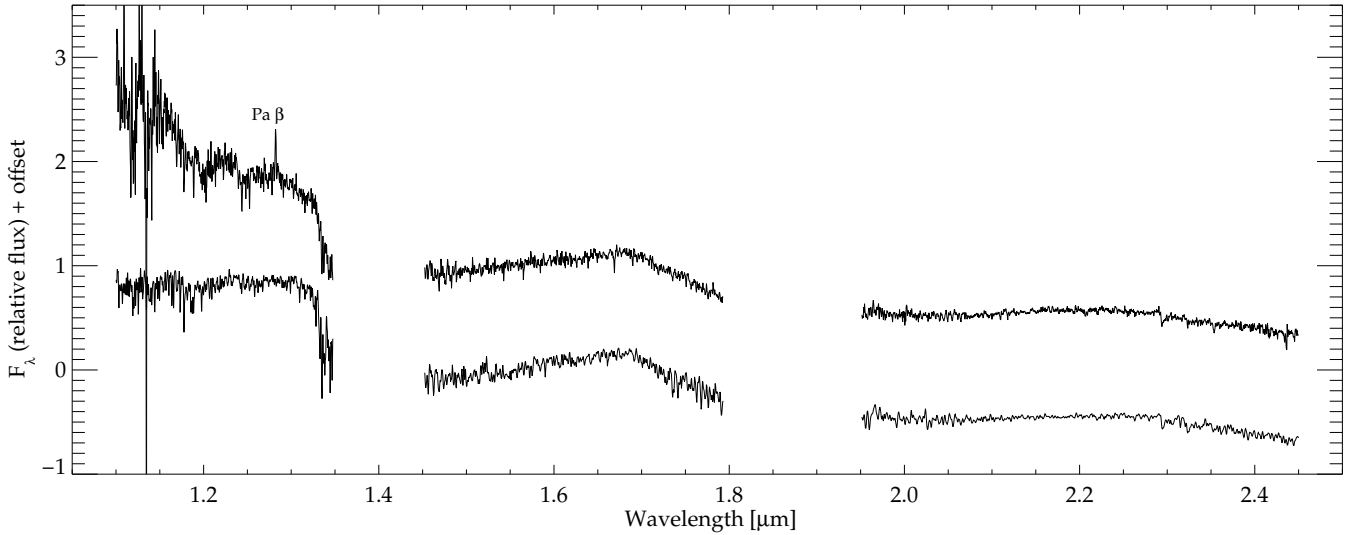}}
   \caption{J, H, and K band spectra, adjusted in their relative flux levels. \textit{From top to bottom:} Our SINFONI spectra of the CT Cha companion in spectral resolution 1500 in comparison to the best-fitting Drift-Phoenix synthetic spectrum (same spectral resolution) of T$_{\rm eff}$=\,2600\,K, $\log{g}$=\,3.5, [Me/H]=\,0.0 and a visual extinction of A$_{V}$=\,5.2 mag. Note the strong Pa-$\beta$ emission line at $\sim$\,1.282 $\mu$m in the J-band and the weak alkali lines of \ion{K}{I} at $\sim$\,1.25 $\mu$m and of \ion{Al}{I} at $\sim$\,1.315 $\mu$m) in the J-band of the synthetic model. See also the online material for a colored version.}
   \label{spectrum_CTCha}
   \end{figure*}

Due to the high number of free parameters after taking the metallicity and the local extinction in the Chamaeleon region into account, which can be up to $A_{V}$\,=\,18\,mag \citep{2006A&A...447..597K} depending on the position in the Chamaeleon star-forming region, we used a $\chi^2$ minimization algorithm to find the best-fitting combination of (a) the effective temperature, (b) the surface gravity, (c) the metallicity, and (d) the extinction correction of the measured spectrum. To determine the extinction, we assumed that a possible circumstellar disk makes a negligible contribution to our near-IR spectra, a reasonable assumption given that brown dwarf disks are too cool to provide any significant near-IR excess emission \citep{2003ApJ...585..372L,2001A&A...376L..22N,2007ApJ...657..511A}. As described in \cite{2007A&A...463..309S}, uncertainties from the noise in the spectra were found to be insignificant compared to the intrinsic uncertainties in the synthetic spectra.

We find a best fit for the companion candidate of CT Cha at T$_{\rm eff}$=\,2600\,$\pm$\,250\,K, a visual extinction $A_{V}$=\,5.2\,$\pm$\,0.8\,mag,  $\log{g}$=\,3.5, and [Me/H]=\,0.0, shown in Fig.~\ref{spectrum_CTCha}.
All values are derived from our $\chi^2$ minimization analysis, see also Fig.\,\ref{chi_CTCha}. While the error bars of the temperature are quite high due to high degeneracy of effective temperature and extinction, the error bars of the surface gravity and metallicity are both $\geq$\,0.5 dex, hence giving no significant constraints for these parameters.



Several signs of youth are present in the spectrum of the CT Cha companion, like the depth of the KI lines in J-band, the perfectly triangular shaped H-band, which lacks any sign of the FeH absorption bands usually seen in old field dwarfs, or the 
slope of the bluest part of the J band as already found for the young companion of GQ Lup \citep{2007A&A...463..309S} and for 2MASS J01415823-4633574 \citep{2006ApJ...639.1120K}.

\subsection{Comparison of {\sc Drift-Phoenix} to \newline USco J160648-223040}

In Fig.~\ref{spectrum_USco J160648-223040} we compare the spectrum of the young free-floating object USco J160648-223040 to the best-fitting model of its spectrum from the same {\sc Drift-Phoenix} model grid used in the previous section.
This free-floating brown dwarf in the $\sim$\,5\,Myr old Upper Scorpius association was found in an infrared photometric survey \citep{2007MNRAS.379.1599L} and later classified with the (optical) spectral type M8 \citep{2008MNRAS.383.1385L} by perfectly matching the infrared spectrum of SCH 162528.62-165850.55, a known member classified optically as M8 \citep{2006AJ....131.3016S}.

\begin{figure}
   \resizebox{\hsize}{!}{\includegraphics{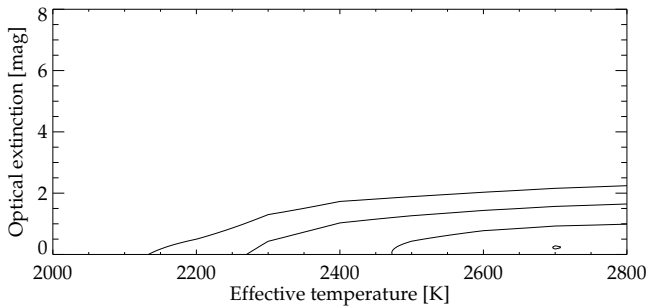}}
   \caption{Result of the $\chi^2$ minimization analysis for USco J160648-223040. Plotted are the best value (\textit{point}) and the 1, 2, and 3 sigma error contours for effective temperature T$_{\rm eff}$ and optical extinction $A_{V}$, determined from comparison of the spectrum of USco J160648-223040 and the Drift-Phoenix model grid.}
   \label{chi_UScoJ160648-223040}
   \end{figure}

\begin{figure*}
   \resizebox{\hsize}{!}{\includegraphics{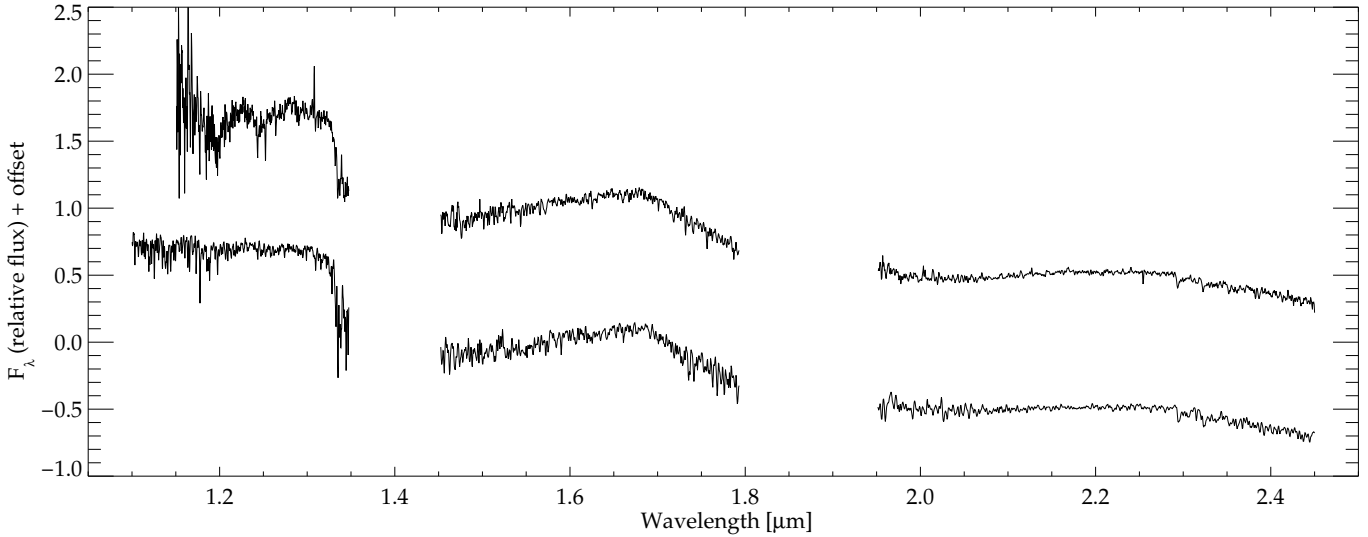}}
   \caption{J, H, and K band spectra, adjusted in their relative flux levels. \textit{From top to bottom:} The spectrum of the free-floating brown dwarf USco J160648-223040 \citep{2008MNRAS.383.1385L} in spectral resolution 1500 in comparison to the best-fitting Drift-Phoenix synthetic spectrum (same spectral resolution) of T$_{\rm eff}$=\,2700\,K, $\log{g}$=\,3.5, [Me/H]=\,-0.5 and a visual extinction of A$_{V}$=\,0.2 mag. See also the online material for a colored version.}
   \label{spectrum_USco J160648-223040}
   \end{figure*}

We find a best fit (Figs.~\ref{chi_UScoJ160648-223040}, \ref{spectrum_USco J160648-223040}) for USco J160648-223040 using the {\sc Drift-Phoenix} models T$_{\rm eff}$= \,2700\,$\pm$\,250\,K, $A_{V}$=\,0.2\,$^{+\,0.8}_{-\,0.2}$\,mag at $\log{g}$=\,3.5 and [Me/H]=\,-0.5. The derived best-fitting temperature of 2700\,K nicely matches both values of 2720\,K and 2710\,K using the former and new temperature (to spectral type conversion) scale of \citet{1999ApJ...525..466L} and \citet{2003ApJ...593.1093L}, respectively, for a(n optical) spectral type of M8 and for objects intermediate between dwarfs and giants, appropriate for young, not fully contracted, sub-stellar objects.

Recently, \citet{2007ApJ...657..511A} have tested several H$_{2}$O indices from the literature and find that a spectral index defined as $\langle F_{\lambda=1.550-1.560} \rangle$ / $\langle F_{\lambda=1.492-1.502} \rangle$ and used with low-resolution spectra yields an index-SpT relationship that is independent of gravity and shows a linear relation between the index and the spectral type. We used this index to determine a spectral type for USco J160648-223040 independent of {\sc Drift-Phoenix} models. From the mid-resolution spectrum of USco J160648-223040, we find the spectral index to be $\sim$\,1.0584
originally and $\sim$\,1.0558 after dereddening $A_{V}$=\,0.2\,mag, as found by our best fit using the synthetic atmospheric models. If we then use Eq.~(1) from \citet{2007ApJ...657..511A} (fit for field dwarfs, giants, and young standards for spectral types M5 -- L0) to invert this index to a spectral type, we get M7.01 and M6.95, respectively, hence M7.

This may be a discrepancy from the temperature scale in \citet{2003ApJ...593.1093L}, consistent with the spectral type determined from spectral comparison to an optical M8 standard and to our derived temperature of $\sim$2700\,K.
As mentioned in \citet{2008arXiv0805.3722L}, the uncertainties in the temperature scale are at least $\pm$\,100\,K, as in the case of our derived temperatures, so that the difference is still within the derived errors.
However, the spectral types in \citet{2007ApJ...657..511A} are near-IR determined spectral types. As argued by \citet{2006PASP..118..611G} and \citet{2005ARA&A..43..195K}, it is neither uncommon nor unexpected that optical and near-IR spectral types do not agree for old L- and T-dwarfs and can even differ for M-dwarfs \citep{2008Meyer}. The closest example we could find to USco J160648-223040 is 2MASS J1204303+321259, which was classified optically as L0 by \citet{2003AJ....126.2421C} and in the near-IR as M9 by \citet{2003IAUS..211..197W}.
We conclude that the most probable solution to this discrepancy is that the near-IR spectral type of M7 of USco J160648-223040 corresponds to an optical spectral type of $\sim$\,M8.
We can, then, further conclude that the companion candidate to CT Cha has a spectral type $\geq$\,M8, as the best fit for its effective temperature of T$_{\rm eff}$=\,2600\,K is 100\,K below the best fit of T$_{\rm eff}$=\,2700\,K for USco J160648-223040, which corresponds to spectral type M8 of the used optical comparison standard via the temperature scale \citep{2003ApJ...593.1093L}.

\subsection{Comparison to USco J160714-232101}

\begin{figure}
   \resizebox{\hsize}{!}{\includegraphics{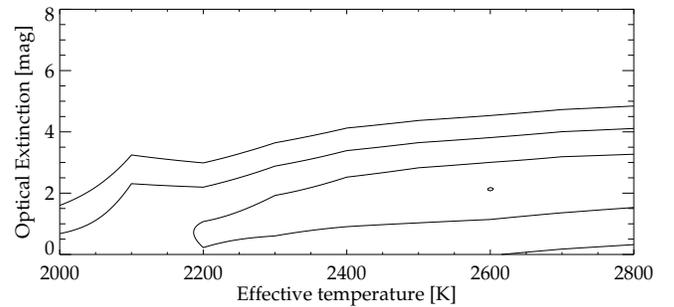}}
   \caption{Result of the $\chi^2$ minimization analysis for USco J160714-232101. Plotted are the best value (\textit{point}) and the 1, 2, and 3 sigma error contours for effective temperature T$_{\rm eff}$ and optical extinction $A_{V}$, determined from comparison of the spectrum of USco J160714-232101 and the Drift-Phoenix model grid.}
   \label{chi_USco J160714-232101}
   \end{figure}

\begin{figure*}
   \resizebox{\hsize}{!}{\includegraphics{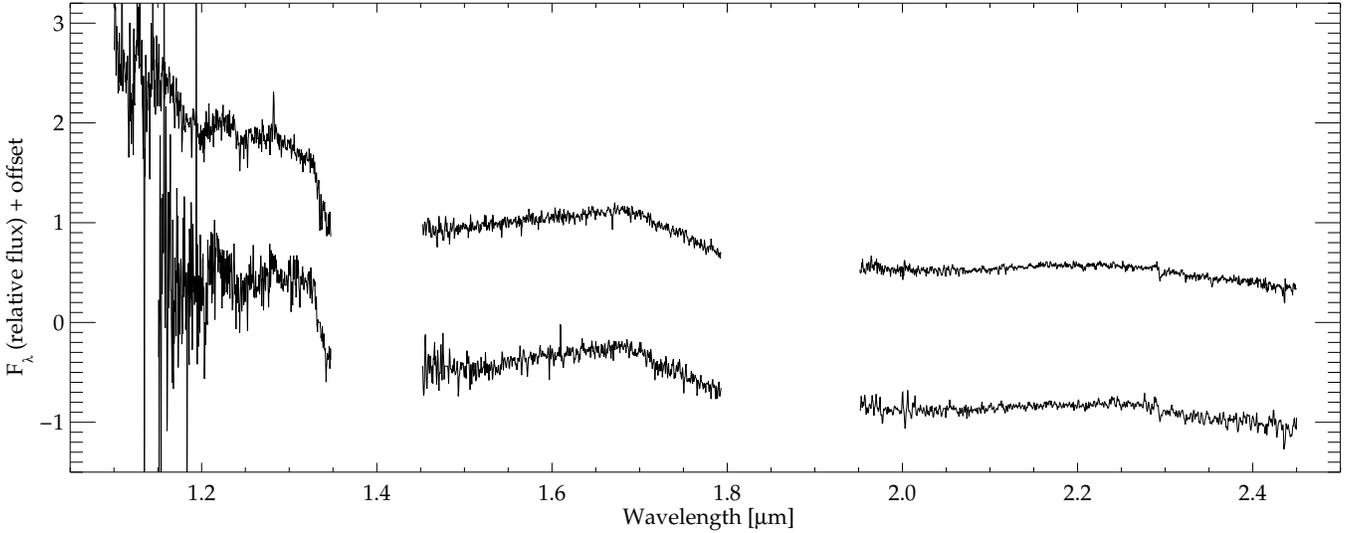}}
   \caption{J, H, and K band spectra, adjusted in their relative flux levels. \textit{From top to bottom:} Our SINFONI spectra of the CT Cha companion in spectral resolution 1500 in comparison to the spectrum of the free-floating brown dwarf USco J160714-232101 \citep{2008MNRAS.383.1385L} (same spectral resolution). The fit of both objects with Drift-Phoenix gives T$_{\rm eff}$=\,2600\,K, $\log{g}$=\,3.5, [Me/H]=\,0.0, while the CT Cha companion and USco J160714-232101 have A$_{V}$=\,5.2\,mag and A$_{V}$=\,2.1\,mag, respectively. See the online material for a colored version.}
   \label{spectrum_USco J160714-232101}
   \end{figure*}

In Fig.~\ref{spectrum_USco J160714-232101} we compare our SINFONI spectra of the CT Cha companion with another young free-floating brown dwarf from the $\sim$\,5\,Myr old Upper Scorpius association USco J160714-232101 \citep{2008MNRAS.383.1385L}.
For this object we find a best fit in comparison to the {\sc Drift-Phoenix} grid, see Fig.~\ref{chi_USco J160714-232101}, for T$_{\rm eff}$=\,2600\,$\pm$\,400\,K, $A_{V}$=\,2.1\,$\pm$\,1.9\,mag at $\log{g}$=\,3.5 and Me/H]=\,0.0.

As can be seen in Fig.~\ref{spectrum_USco J160714-232101}, our SINFONI spectra of the CT Cha companion and the spectrum of USco J160714-232101 match very well after dereddening.
USco J160714-232101 is given as a spectral type L0 member of the Upper Scorpius association in \citet{2008MNRAS.383.1385L} from comparison to the infrared spectra of two standards of spectral types M8 and M9, which were characterized by their optical spectra, and neglecting possible present extinction. We had to deredden the spectrum of USco J160714-232101 by A$_{V}$=\,2.1\,mag to match the {\sc Drift-Phoenix} atmospheric models, so we conclude that the CT Cha companion has a spectral type of $\leq$ L0 and, in combination with the result from the previous section, an optical spectral type of M8 -- L0.

\subsection{Comparison to CHXR 73 B}

\citet{2006ApJ...649..894L} present the discovery of a 
similar object, which is interesting as a comparison object, as it is not member of a slightly older association as the already presented Upper Scorpius objects, but also a member of the very young Cha I star-forming region. They conclude that the object is very likely a companion of CHXR 73 A and that it has a mass of 0.012$^{+ 0.008}_{- 0.005}$ M$_{\odot}$.

In Fig.~\ref{chi_CHXR73} we show the $\chi^2$ minimization results for this object giving a best fit in comparison to the {\sc Drift-Phoenix} grid for T$_{\rm eff}$=\,2600\,$\pm$\,450\,K, $A_{V}$=\,12.6\,$\pm$\,2.1\,mag at $\log{g}$=\,4.5 and [Me/H]=\,0.0. The higher uncertainties in temperature and visual extinction at the same temperature 
as for the CT companion candidate are due to high reddening of CHXR 73 B resulting in a low signal-to-noise (S/N) and high correlation of temperature and extinction. The accordingly dereddened spectrum of CHXR 73 B is compared to the companion of CT Cha in Fig.~\ref{spectrum_CHXR73}.

\begin{figure}
   \resizebox{\hsize}{!}{\includegraphics{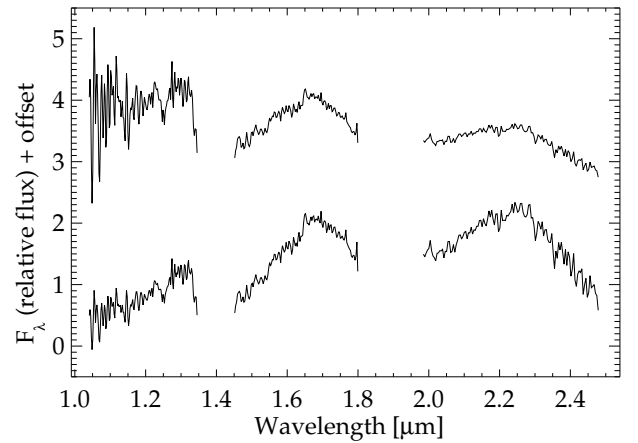}}
   \caption{Near-IR J, H, and K band spectra of CHXR 73 B from \citet{2006ApJ...649..894L}. \textit{From top to bottom:} Spectrum dereddened by A$_{V}$=\,7.6\,mag as given as A$_{J}\sim$\,2.1\,mag \citep{2006ApJ...649..894L} and transformed using the extinction law of \citet{1985ApJ...288..618R}. Reddened spectrum as originally measured.}
   \label{Dered_CHXR73B}
   \end{figure}

\begin{figure}
   \resizebox{\hsize}{!}{\includegraphics{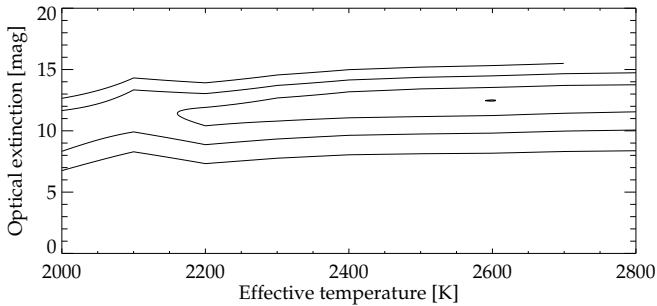}}
   \caption{Result of the $\chi^2$ minimization analysis for CHXR 73 B. Plotted are the best value (\textit{point}) and the 1, 2, and 3 sigma error contours for effective temperature T$_{\rm eff}$ and optical extinction $A_{V}$, determined from comparison of the spectrum of CHXR 73 B and the Drift-Phoenix model grid convolved to the same resolution.}
   \label{chi_CHXR73}
   \end{figure}

\begin{figure*}
   \resizebox{\hsize}{!}{\includegraphics{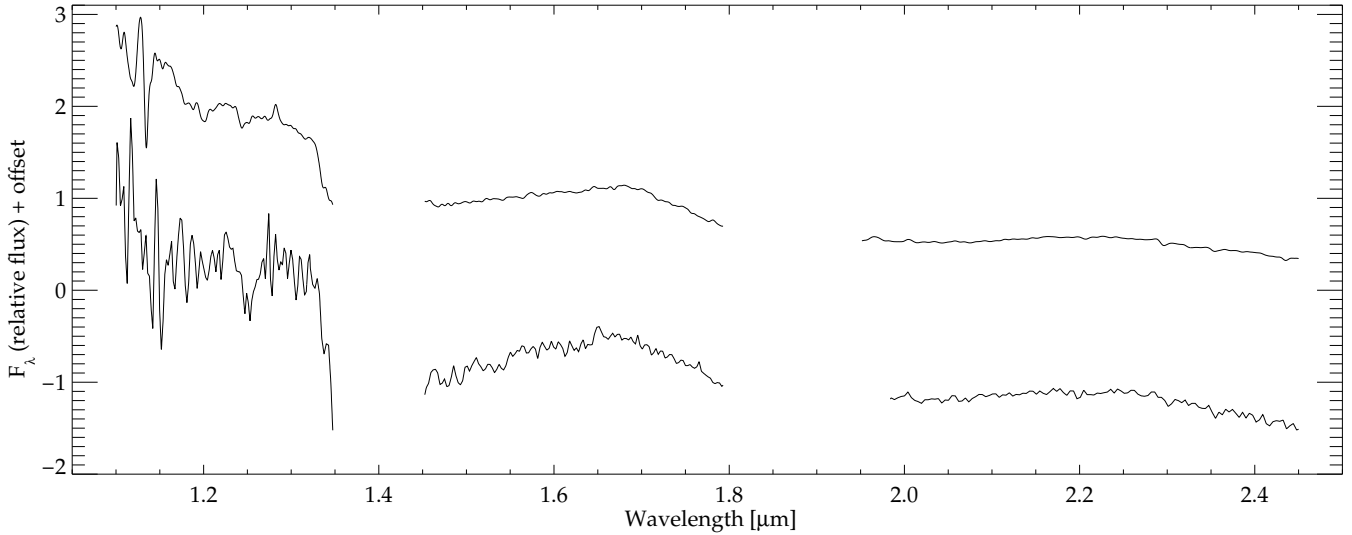}}
   \caption{J, H, and K band spectra, adjusted in their relative flux levels. \textit{From top to bottom:} Our SINFONI spectra of the CT Cha companion in spectral resolution 300 in comparison to the spectrum of the companion CHXR 73 B  \citet{2006ApJ...649..894L}, as well member of Cha I (same spectral resolution). The fit of both objects with Drift-Phoenix gives 2600 K, [Me/H]=\,0.0, while the CT companion and CHXR 73 B have A$_{V}$=\,5.2\,mag, $\log{g}$=\,3.5, and A$_{V}$=\,12.6\,mag,  $\log{g}$=\,4.5  respectively. See also the online material for a colored version.}
   \label{spectrum_CHXR73}
   \end{figure*}

As can be seen from this comparison, there are deviant slopes in the blue part of the H-band and the red part of the K-band. However, these parts of the spectrum have a low S/N due to the high extinction in comparison to the rest of the corresponding bands, as can be seen in Fig.~\ref{Dered_CHXR73B}. We find an unusually high surface gravity, $\log{g}$=\,4.5 being at the edge of our extended {\sc Drift-Phoenix} model grid at this temperature. This is supported by the deep potassium lines at $\sim$\,1.25\,$\mu$m, however, the whole J-band also has a low S/N as already dicussed in \citet{2006ApJ...649..894L}.

Finally the new extinction value of $A_{V}$=\,12.6\,$\pm$\,2.1\,mag is much higher than the originally derived value of $A_{V}\sim$\,7.6\,mag, converted from $A_{J}\sim$\,2.1\,mag in \citet{2006ApJ...649..894L} using the extinction law by \citet{1985ApJ...288..618R}. This can be partly explained by the extinction originally having been determined in comparison to KPNO-Tau 4 \citep[$A_{J}\sim$\,0\,mag][]{2002ApJ...580..317B} in \citet{2006ApJ...649..894L}, while \citet{2007A&A...465..855G} give a value of $A_{V}$=\,2.45\,mag. The remaining discrepancy can probably be attributed to measurement uncertainties, as the uncertainty we found of $\pm$\,2.1\,mag already covers 82\,\% of the residual difference to our new derived value of $A_{V}$=\,12.6\,$\pm$\,2.1\,mag.

If we assume the new extinction value of $A_{V}$=\,12.6\,$\pm$\,2.1\,mag to be correct, the luminosity of CHXR 73 B changes from $\log{L_{bol}/L_{\odot}}$= -2.85\,$\pm$\,0.14 \citep{2006ApJ...649..894L} to $\log{L_{bol}/L_{\odot}}$= \mbox{-2.56}\,$\pm$\,0.22 if we assume in addition B.C.$_{K}$=\,3.15\,$\pm$\,0.15\,mag (corresponding to an optical spectral type of M8 -- L0 as for the CT Cha companion), as well as the distance of 165\,$\pm$\,30\,pc and an absolute bolometric magnitude of the sun M$_{bol\odot}$=\,4.74\,mag.

Although there is no final proof that the nature of the CT Cha companion and CHXR 73 B are the same, most indications (e.g. luminosity, temperature, hence spectral type) point to this explanation. Moreover, the spectra match very well in the areas of good S/N after correction for extinction. However, we cannot exclude from this comparison that the deviant slopes in the blue part of the H-band and the red part of the K-band, as well as the deep potassium lines at $\sim$\,1.25\,$\mu$m, are indeed real and connected to a higher age and surface gravity for CHXR 73 B.

\subsection{Comparison to 2MASS J01415823-4633574}

We also compared our SINFONI spectra of the CT Cha companion with the young free-floating object 2MASS J01415823-4633574 found by \citet{2006ApJ...639.1120K}.
Our best fit for this object in comparison to the {\sc Drift-Phoenix} grid suggests the same temperature of 2600\,K as for the CT Cha companion and $\geq$\,4\,mag of extinction. However, only the H- and K-band show a good fit in this case. Slightly more reddening in the J-band would solve this problem for the J-band but would decrease the quality of the fit in the other two bands, pointing however towards a lower temperature. Moreover, the K I lines at $\sim$\,1.25\,$\mu$m are slighly deeper than in the CT Cha companion, while the gravity sensitive sodium doublet in the K-band cannot be seen in 2MASS J01415823-4633574, pointing towards a slightly higher surface gravity, and we find an unusually deepness of the broad VO feature whose center is at $\sim$\,1.20\,$\mu$m.

If we assume that no extinction can be present, or not much, since the object is lying in the field, not in a young star-forming association, we find an H$_{2}$O index (already used in Sect.~3.3) from the measured H-band spectrum of $\sim$\,1.16856 corresponding to infrared spectral type M9.5 \citep{2007ApJ...657..511A}, while its optical spectral type is probably even later. Since the temperature scale of \citet{2003ApJ...593.1093L} ends at M9, we cannot find an appropriate temperature equivalent for this probably young object.

It is challenging to explain the spectrum of 2MASS J01415823-4633574, even though the {\sc Drift-Phoenix} models match objects of slighly earlier spectral types nicely. The most probable explanations are that 2MASS J01415823-4633574 might either have properties lying outside of the values used in our grid or that it is indeed a peculiar object as discussed in \citet{2006ApJ...639.1120K}, hence we suggest keeping the spectral type L0pec for now.
Generally, we find a good comparison for non-peculiar objects with the {\sc Drift-Phoenix} models, which shows that these new model atmospheres are useful.

\section{Mass estimation and discussion}


We derived a luminosity of $\log(L_{bol}/L_{\odot})$=\,-2.68\,$\pm$\,0.21 for the CT Cha companion from the extinction corrected values Ks$_{0}$= 14.37\,$\pm$\,0.32\,mag (2006), J$_{0}$= 15.25\,$\pm$\,0.39\,mag, and Ks$_{0}$= 14.31\,$\pm$\,0.32\,mag (2007) \cite[from our A$_{V}$ and extinction law by][]{1985ApJ...288..618R} using a bolometric correction of B.C.$_{K}$\,=\,3.15\,$\pm$\,0.15\,mag from \citet[][for spectral type M8--L0]{2004AJ....127.3516G} and adopting a distance of 165\,$\pm$\,30\,pc. From the luminosity and temperature, we calculated the radius to be R=\,0.23$_{-0.06}^{+0.08}$\,R$_{\odot}$ or 2.20$_{-0.60}^{+0.81}$\,R$_{\mathrm{Jup}}$.

We considered several possibilities for deriving a final mass. Since our S/N is too low in the very blue region of the J-band, we did not use the gravity-sensitive sodium index from \citet{2007ApJ...657..511A}. We likewise cannot use the depth of the alkali lines, because they are still in the process of being included in the {\sc Drift-Phoenix} atmospheric models, which are the only models we found able to reproduce the J, H, and K continua of the CT Cha companion without any additional assumptions.
As a result, we use evolutionary models to get a mass estimation of the object, stressing, however, that such models are uncertain up to at least $\sim$\,10 Myrs due to unknown initial conditions \citep{2005astro.ph..9798C}. For an age of 2\,$\pm$\,2\,Myr, we find from luminosity and temperature masses of 8 -- 23 M$_{Jup}$, 13.5 -- 60 M$_{Jup}$ \citep{2003A&A...402..701B} and 9 -- 23 M$_{Jup}$, 12.5 -- 50 M$_{Jup}$ \citep{2000ApJ...542..464C} and $\sim$\,9 -- 22 M$_{Jup}$, $\sim$\,11 -- 60 M$_{Jup}$ \citep{1997ApJ...491..856B}, respectively. The best-fitting models are from \citet{1997ApJ...491..856B} and give for an age of 2\,Myr a temperature of 2540\,K, a surface gravity of $\log{g}$= 3.9, and a luminosity of $\log(L_{bol}/L_{\odot})$= -2.64 a mass of 17 M$_{Jup}$. The possible mass intervals from luminosity and temperature of all evolutionary models overlap, so we conclude from the intersection of these intervals that our companion candidate has a mass of 17\,$\pm$\,6\,M$_{Jup}$.

To obtain a limit from direct observational evidence, we can compare the CT Cha companion with the eclipsing double-lined spectroscopic binary brown dwarf 2MASS J05352184–-0546085 found by \citet{2006Natur.440..311S} with known masses from first principles and of a comparable age to CT Cha. From this comparison we conclude that, due to much lower luminosity, lower temperature, and radius, the CT Cha companion must be lower in mass than all the components in 2MASS J05352184–-0546085, i.e.\,$\leq$\,35\,M$_{\mathrm{Jup}}$ \citep{2007ApJ...664.1154S}.


From the comparison to objects in the Upper Scorpius association \citep{2008MNRAS.383.1385L}, we found an optical spectral type of the CT Cha companion of M8 -- L0. We note that our object is very likely a member of the Cha I star-forming region, because it is co-moving with the primary, a member of Cha I, and because the equivalent width of the two combined potassium (K I) lines around 1.25\,$\mu$m of 9\,$\pm$\,5\,\AA \ is smaller, just as for USco J160714-232101, pointing to a lower age than $\sim$\,5\,Myr since this object is a member of the Upper Scorpius association \citep{2008MNRAS.383.1385L}. From our fits to the {\sc Drift-Phoenix} models, we found a likely metallicity of [Me/H]=0.0. Although this value is not very well constrained, it is consistent with the median value $\langle[Fe/H]\rangle$ of -0.11 found for some members of Chamaeleon \citep{2008A&A...480..889S}.

At a projected distance of $\sim$\,440 AU, the system is probably still below the long-term stability limit of $\sim$\,700 AU for a K7 primary star of 0.7\,M$_{\odot}$, following the argumentation of \citet{1987ApJ...312..367W} and \citet{2003ApJ...587..407C} \citep[see also][for a discussion]{2005AN....326..701M}.

Brown dwarf companions of slightly higher mass have also been found as companions to much older stars and at a variety of projected separations. Examples are GJ 229 B \citep{1995Natur.378..463N,1995Sci...270.1478O} at a projected separation of $\sim$\,45\,AU and the binary brown dwarf companion Eps Indi B \citep{2003A&A...398L..29S} at a projected separation of $\sim$\,1460\,AU from their primaries. The currently most similar co-moving system to CT Cha A and its companion at a higher age is, however, HD 3651 B \citep{2006MNRAS.373L..31M} and its host star found to have a projected separation of $\sim$\,480\,AU.

The prominent Pa-$\beta$ emission line in the J-band is most likely due to ongoing accretion onto the CT Cha companion, produced in the shock-heated magnetospheric accretion flow \citep{1998AJ....116..455M,2004A&A...417..247W,2005MNRAS.358..671K} and thus another sign of its youth.
While this magnetospheric accretion model predicts a correlation between emission line strengths of hydrogen lines and accretion rate \citep{1998AJ....116..455M}, as used to determine accretion rates from the H-$\alpha$ emission line, \citet{2001A&A...365...90F} find no such agreement for the Pa-$\beta$ emission line of approximately half of the 49 studied classical T-Tauri stars \citep[see][for a discussion]{2004A&A...417..247W}.
Interestingly, we did not find any emission in the Br-$\gamma$ line around 2.166 \AA . While a possible emission of Br-$\gamma$ might be present below the detection limit, the same behavior, having only Pa-$\beta$ emission, was also found for the companion of GQ Lup \citep{2007A&A...463..309S}.

Very interesting is the high extinction value of $A_{V}$=\,5.2\,$\pm$\,0.8\,mag of the CT Cha companion compared to the extinction value of CT Cha A, which we determined with the J-, H-, and K-band 2MASS photometry partly given in Sect.~2.3 and in comparison to the colors of main-sequence stars given in \citet{1995ApJS..101..117K}. We find a best fit for $A_{V}\sim$\,1.3\,mag and spectral type K7 consistent with the spectral type K7 by \citet{1988A&AS...76..347G}. This is in good agreement with $A_{V}\sim$\,1.4\,mag found by \citet{1998A&A...338..977C} from I\,-\,J color excess, but inconsistent with the $A_{V}\sim$\,0.1\,mag found by the same authors using star counts.

This behavior is very similar for CHXR 73 A and its companion CHXR 73 B, which has 
similar properties as the CT Cha companion and thus both are probably similar in nature. We found an extinction of CHXR 73 B of $A_{V}$=\,12.6\,$\pm$\,2.1\,mag from fitting the spectrum of CHXR 73 B to {\sc Drift-Phoenix} models, afterwards matching in the highest signal-to-noise areas the spectra of the CT Cha companion. We find the same best temperature 2600\,K fitting both objects, while the luminosities are consistent within their 1\,$\sigma$ error bars.
Even the difference of primary to secondary optical extinction is similar, being 3.1 -- 4.7\,mag for CT Cha and 1.7 -- 5.9\,mag for CHXR 73, derived from comparison of our CHXR 73 B value to the value of CHXR 73 A $A_{J}$=\,1.8\,mag \citep{2004ApJ...602..816L} and using the extinction of KPNO-Tau 4 $A_{V}$=\,2.45\,mag \citep{2007A&A...465..855G} (see Sect.~3.5).
However, we can still not exclude from our spectral comparison that the deviant slopes in the blue part of the H-band and the red part of the K-band (see Fig.~\ref{spectrum_CHXR73}), as well as the deep potassium lines at $\sim$\,1.25\,$\mu$m of CHXR 73 B, are indeed real and connected to a higher age and surface gravity of CHXR 73 B.

This difference in the extinction of the objects can have any of the following three causes. (1) The extinction in the young Chamaeleon I star-forming region is very clumpy and changes drastically on short spatial scales. (2) Although a close to pole-on  disk around a sub-stellar object would make no significant contribution to the spectrum in the near-infrared, an object bearing a close to edge-on disk could be reddened in this spectral region due to e.g. reflection and absorption by the disk. (3) The companions of CHXR 73 and CT Cha are both a projection effect and are actually situated further to the back of the Cha I cloud, still co-moving as they are members of the star-forming region, sharing the same space motion in a moving group, but not being physically bound to their primary objects. This can hardly be reconciled, however, with the probabilities of finding a by chance alignment estimated to be $\sim$\,0.001 for CHXR 73 B by \citet{2006ApJ...649..894L} and $\sim$\,0.01 for the CT Cha companion by us (see Sect.~2.2).

We conclude from our mass estimate of 17\,$\pm$\,6\,M$_{Jup}$ that the CT Cha companion is most likely a brown dwarf, we can, however, not exclude the possibility that the CT Cha companion is a planetary mass object. There is yet no consensus or definition for \textit{planets} around other stars, or for the upper mass limit of planets. In general we can follow the discussion in \citet{2006ApJ...649..894L} saying that CHXR 73 B was probably not formed in a planetary fashion as its separation from the primary is very high, as is the case for the CT Cha companion, and cloud fragmentation was shown to be able to produce even isolated objects of similar masses. However we would like to stress that this argumentation assumes that the companions actually formed at the current separation to their primaries. We cannot yet rule out that the objects were scattered outward, as supposed for the sub-stellar companion of GQ Lup by \citet{2006ApJ...637L.137B} and \citet{2006A&A...451..351D} and supported by planet-planet interaction models as e.g. from \citet{2008ApJ...678..498N}, where planets can be scattered to highly excentric orbits and large semi-major axis by interaction. 

New evolutionary models by \citet{2007ApJ...655..541M} show that objects created by core accretion have lower luminosities at very young ages than expected before. As CT Cha A and its companion have an estimated age of 2\,$\pm$\,2\,Myr, it is thus unlikely that the companion was formed by core accretion.




\begin{acknowledgements}
TOBS acknowledges support from the Evangelisches Studienwerk e.V. Villigst.
Moreover, he thanks his family for constant support.

NV acknowledges support by FONDECYT grant 1061199.

AB and RN would like to thank the DFG for financial support in projects NE 515 / 13-1 and 13-2.

In addition we would like to thank T. Tsuji for his synthetic atmospheric models, especially the calculations of the UCM-models for lower surface gravities, and for his encouraging comments to this work.

Furthermore, we thank J. D. Kirkpatrick for the electronic version of the spectrum of 2MASS J01415823-4633574, K. Luhman for the electronic version of the spectrum of CHXR 73 B, N. Lodieu for the electronic versions of the spectra of USco J160648-223040 and USco J160714-232101, the ESO staff of the VLT at Paranal for their support and the execution of the service mode observations, M.\,R.~Meyer and an anonymous referee for helpful comments and Joli Adams for language editing.

This publication makes use of data products from the Two Micron All Sky Survey, which is a joint project of the University of Massachusetts and the Infrared Processing and Analysis Center/California Institute of Technology, funded by the National Aeronautics and Space Administration and the National Science Foundation. This work makes use of a conversion tool developed by the Spitzer Science Center, which hosts the Spitzer Space Telescope operated by the Jet Propulsion Laboratory, California Institute of Technology under a contract with NASA. 

This research has made use of the VizieR catalog access tool and the Simbad database, both operated at the Observatoire Strasbourg
and has benefited from the M, L, and T dwarf compendium housed at DwarfArchives.org and maintained by Chris Gelino, Davy Kirkpatrick, and Adam Burgasser. The NSO/Kitt Peak FTS data used here were produced by NSF/NOAO.

\end{acknowledgements}

\bibliographystyle{aa}

\begin{thebibliography}{94}
\expandafter\ifx\csname natexlab\endcsname\relax\def\natexlab#1{#1}\fi

\bibitem[{{Abuter} {et~al.}(2006){Abuter}, {Schreiber}, {Eisenhauer}, {Ott},
  {Horrobin}, \& {Gillesen}}]{2006NewAR..50..398A}
{Abuter}, R., {Schreiber}, J., {Eisenhauer}, F., {et~al.} 2006, New Astronomy
  Review, 50, 398

\bibitem[{{Ackerman} \& {Marley}(2001)}]{2001ApJ...556..872A}
{Ackerman}, A.~S. \& {Marley}, M.~S. 2001, \apj, 556, 872

\bibitem[{{Allard} {et~al.}(2007){Allard}, {Allard}, {Homeier}, {Kielkopf},
  {McCaughrean}, \& {Spiegelman}}]{2007A&A...474L..21A}
{Allard}, F., {Allard}, N.~F., {Homeier}, D., {et~al.} 2007, \aap, 474, L21

\bibitem[{{Allers} {et~al.}(2007){Allers}, {Jaffe}, {Luhman}, {Liu}, {Wilson},
  {Skrutskie}, {Nelson}, {Peterson}, {Smith}, \&
  {Cushing}}]{2007ApJ...657..511A}
{Allers}, K.~N., {Jaffe}, D.~T., {Luhman}, K.~L., {et~al.} 2007, \apj, 657, 511

\bibitem[{{Baraffe} {et~al.}(2003){Baraffe}, {Chabrier}, {Barman}, {Allard}, \&
  {Hauschildt}}]{2003A&A...402..701B}
{Baraffe}, I., {Chabrier}, G., {Barman}, T.~S., {Allard}, F., \& {Hauschildt},
  P.~H. 2003, \aap, 402, 701

\bibitem[{{Batalha} {et~al.}(1998){Batalha}, {Quast}, {Torres}, {Pereira},
  {Terra}, {Jablonski}, {Schiavon}, {de La Reza}, \&
  {Sartori}}]{1998A&AS..128..561B}
{Batalha}, C.~C., {Quast}, G.~R., {Torres}, C.~A.~O., {et~al.} 1998, \aaps,
  128, 561

\bibitem[{{Bertout} {et~al.}(1999){Bertout}, {Robichon}, \&
  {Arenou}}]{1999A&A...352..574B}
{Bertout}, C., {Robichon}, N., \& {Arenou}, F. 1999, \aap, 352, 574

\bibitem[{{Boss}(2006)}]{2006ApJ...637L.137B}
{Boss}, A.~P. 2006, \apjl, 637, L137

\bibitem[{{Brice{\~n}o} {et~al.}(2002){Brice{\~n}o}, {Luhman}, {Hartmann},
  {Stauffer}, \& {Kirkpatrick}}]{2002ApJ...580..317B}
{Brice{\~n}o}, C., {Luhman}, K.~L., {Hartmann}, L., {Stauffer}, J.~R., \&
  {Kirkpatrick}, J.~D. 2002, \apj, 580, 317

\bibitem[{{Brott} \& {Hauschildt}(2005)}]{2005ESASP.576..565B}
{Brott}, I. \& {Hauschildt}, P.~H. 2005, in ESA Special Publication, Vol. 576,
  The Three-Dimensional Universe with Gaia, ed. C.~{Turon}, K.~S. {O'Flaherty},
  \& M.~A.~C. {Perryman}, 565

\bibitem[{{Burrows} {et~al.}(1997){Burrows}, {Marley}, {Hubbard}, {Lunine},
  {Guillot}, {Saumon}, {Freedman}, {Sudarsky}, \&
  {Sharp}}]{1997ApJ...491..856B}
{Burrows}, A., {Marley}, M., {Hubbard}, W.~B., {et~al.} 1997, \apj, 491, 856

\bibitem[{{Burrows} {et~al.}(2006){Burrows}, {Sudarsky}, \&
  {Hubeny}}]{2006ApJ...640.1063B}
{Burrows}, A., {Sudarsky}, D., \& {Hubeny}, I. 2006, \apj, 640, 1063

\bibitem[{{Camargo} {et~al.}(2003){Camargo}, {Ducourant}, {Teixeira}, {Le
  Campion}, {Rapaport}, \& {Benevides-Soares}}]{Camargo2003}
{Camargo}, J.~I.~B., {Ducourant}, C., {Teixeira}, R., {et~al.} 2003, \aap, 409,
  361

\bibitem[{{Cambresy} {et~al.}(1998){Cambresy}, {Copet}, {Epchtein}, {de Batz},
  {Borsenberger}, {Fouque}, {Kimeswenger}, \& {Tiphene}}]{1998A&A...338..977C}
{Cambresy}, L., {Copet}, E., {Epchtein}, N., {et~al.} 1998, \aap, 338, 977

\bibitem[{{Chabrier} {et~al.}(2000){Chabrier}, {Baraffe}, {Allard}, \&
  {Hauschildt}}]{2000ApJ...542..464C}
{Chabrier}, G., {Baraffe}, I., {Allard}, F., \& {Hauschildt}, P. 2000, \apj,
  542, 464

\bibitem[{{Chabrier} {et~al.}(2005){Chabrier}, {Baraffe}, {Allard}, \&
  {Hauschildt}}]{2005astro.ph..9798C}
{Chabrier}, G., {Baraffe}, I., {Allard}, F., \& {Hauschildt}, P.~H. 2005, ArXiv
  Astrophysics e-prints, astro-ph/0509798

\bibitem[{{Chappelle} {et~al.}(2005){Chappelle}, {Pinfield}, {Steele},
  {Dobbie}, \& {Magazz{\`u}}}]{2005MNRAS.361.1323C}
{Chappelle}, R.~J., {Pinfield}, D.~J., {Steele}, I.~A., {Dobbie}, P.~D., \&
  {Magazz{\`u}}, A. 2005, \mnras, 361, 1323

\bibitem[{{Close} {et~al.}(2003){Close}, {Siegler}, {Freed}, \&
  {Biller}}]{2003ApJ...587..407C}
{Close}, L.~M., {Siegler}, N., {Freed}, M., \& {Biller}, B. 2003, \apj, 587,
  407

\bibitem[{{Cohen} {et~al.}(2003){Cohen}, {Wheaton}, \&
  {Megeath}}]{2003AJ....126.1090C}
{Cohen}, M., {Wheaton}, W.~A., \& {Megeath}, S.~T. 2003, \aj, 126, 1090

\bibitem[{{Cruz} {et~al.}(2003){Cruz}, {Reid}, {Liebert}, {Kirkpatrick}, \&
  {Lowrance}}]{2003AJ....126.2421C}
{Cruz}, K.~L., {Reid}, I.~N., {Liebert}, J., {Kirkpatrick}, J.~D., \&
  {Lowrance}, P.~J. 2003, \aj, 126, 2421

\bibitem[{{Debes} \& {Sigurdsson}(2006)}]{2006A&A...451..351D}
{Debes}, J.~H. \& {Sigurdsson}, S. 2006, \aap, 451, 351

\bibitem[{{Dehn} {et~al.}(2007){Dehn}, {Helling}, {Woitke}, \&
  {Hauschildt}}]{2007IAUS..239..227D}
{Dehn}, M., {Helling}, Ch., {Woitke}, P., \& {Hauschildt}, P. 2007, in IAU
  Symposium, Vol. 239, IAU Symposium, ed. F.~{Kupka}, I.~{Roxburgh}, \&
  K.~{Chan}, 227--229

\bibitem[{{Diolaiti} {et~al.}(2000){Diolaiti}, {Bendinelli}, {Bonaccini},
  {Close}, {Currie}, \& {Parmeggiani}}]{2000SPIE.4007..879D}
{Diolaiti}, E., {Bendinelli}, O., {Bonaccini}, D., {et~al.} 2000, in Presented
  at the Society of Photo-Optical Instrumentation Engineers (SPIE) Conference,
  Vol. 4007, Proc. SPIE Vol. 4007, p. 879-888, Adaptive Optical Systems
  Technology, Peter L. Wizinowich; Ed., ed. P.~L. {Wizinowich}, 879--888

\bibitem[{{Ducourant} {et~al.}(2005){Ducourant}, {Teixeira}, {P{\'e}ri{\'e}},
  {Lecampion}, {Guibert}, \& {Sartori}}]{2005A&A...438..769D}
{Ducourant}, C., {Teixeira}, R., {P{\'e}ri{\'e}}, J.~P., {et~al.} 2005, \aap,
  438, 769

\bibitem[{{Feigelson} {et~al.}(1993){Feigelson}, {Casanova}, {Montmerle}, \&
  {Guibert}}]{1993ApJ...416..623F}
{Feigelson}, E.~D., {Casanova}, S., {Montmerle}, T., \& {Guibert}, J. 1993,
  \apj, 416, 623

\bibitem[{{Folha} \& {Emerson}(2001)}]{2001A&A...365...90F}
{Folha}, D.~F.~M. \& {Emerson}, J.~P. 2001, \aap, 365, 90

\bibitem[{{Gauvin} \& {Strom}(1992)}]{1992ApJ...385..217G}
{Gauvin}, L.~S. \& {Strom}, K.~M. 1992, \apj, 385, 217

\bibitem[{{Gelino} {et~al.}(2006){Gelino}, {Kulkarni}, \&
  {Stephens}}]{2006PASP..118..611G}
{Gelino}, C.~R., {Kulkarni}, S.~R., \& {Stephens}, D.~C. 2006, \pasp, 118, 611

\bibitem[{{Ghez} {et~al.}(1997){Ghez}, {McCarthy}, {Patience}, \&
  {Beck}}]{1997ApJ...481..378G}
{Ghez}, A.~M., {McCarthy}, D.~W., {Patience}, J.~L., \& {Beck}, T.~L. 1997,
  \apj, 481, 378

\bibitem[{{Glass}(1979)}]{1979MNRAS.187..305G}
{Glass}, I.~S. 1979, \mnras, 187, 305

\bibitem[{{Golimowski} {et~al.}(2004){Golimowski}, {Leggett}, {Marley}, {Fan},
  {Geballe}, {Knapp}, {Vrba}, {Henden}, {Luginbuhl}, {Guetter}, {Munn},
  {Canzian}, {Zheng}, {Tsvetanov}, {Chiu}, {Glazebrook}, {Hoversten},
  {Schneider}, \& {Brinkmann}}]{2004AJ....127.3516G}
{Golimowski}, D.~A., {Leggett}, S.~K., {Marley}, M.~S., {et~al.} 2004, \aj,
  127, 3516

\bibitem[{{Goto} {et~al.}(2003){Goto}, {Hayano}, {Kobayashi}, {Terada}, {Pyo},
  {Tokunaga}, {Takami}, {Takato}, {Minowa}, {Gaessler}, \&
  {Iye}}]{2003SPIE.4839.1117G}
{Goto}, M., {Hayano}, Y., {Kobayashi}, N., {et~al.} 2003, in Presented at the
  Society of Photo-Optical Instrumentation Engineers (SPIE) Conference, Vol.
  4839, Adaptive Optical System Technologies II. Edited by Wizinowich, Peter
  L.; Bonaccini, Domenico. Proceedings of the SPIE, Volume 4839, pp. 1117-1123
  (2003)., ed. P.~L. {Wizinowich} \& D.~{Bonaccini}, 1117--1123

\bibitem[{{Gregorio Hetem} {et~al.}(1988){Gregorio Hetem}, {Sanzovo}, \&
  {Lepine}}]{1988A&AS...76..347G}
{Gregorio Hetem}, J.~C., {Sanzovo}, G.~C., \& {Lepine}, J.~R.~D. 1988, \aaps,
  76, 347

\bibitem[{{Guenther} {et~al.}(2007){Guenther}, {Esposito}, {Mundt}, {Covino},
  {Alcal{\'a}}, {Cusano}, \& {Stecklum}}]{2007A&A...467.1147G}
{Guenther}, E.~W., {Esposito}, M., {Mundt}, R., {et~al.} 2007, \aap, 467, 1147

\bibitem[{{Guieu} {et~al.}(2007){Guieu}, {Pinte}, {Monin}, {M{\'e}nard},
  {Fukagawa}, {Padgett}, {Noriega-Crespo}, {Carey}, {Rebull}, {Huard}, \&
  {Guedel}}]{2007A&A...465..855G}
{Guieu}, S., {Pinte}, C., {Monin}, J.-L., {et~al.} 2007, \aap, 465, 855

\bibitem[{{Hartmann} {et~al.}(1998){Hartmann}, {Calvet}, {Gullbring}, \&
  {D'Alessio}}]{1998ApJ...495..385H}
{Hartmann}, L., {Calvet}, N., {Gullbring}, E., \& {D'Alessio}, P. 1998, \apj,
  495, 385

\bibitem[{{Hauschildt} \& {Baron}(1999)}]{1999JCoAM.109...41H}
{Hauschildt}, P.~H. \& {Baron}, E. 1999, Journal of Computational and Applied
  Mathematics, 109, 41

\bibitem[{{Helling} {et~al.}(2008{\natexlab{a}}){Helling}, {Ackerman},
  {Allard}, {Dehn}, {Hauschildt}, {Homeier}, {Lodders}, {Marley}, {Rietmeijer},
  {Tsuji}, \& {Woitke}}]{2008IAUS..249..173H}
{Helling}, Ch., {Ackerman}, A., {Allard}, F., {et~al.} 2008{\natexlab{a}}, in
  IAU Symposium, Vol. 249, IAU Symposium, 173--177

\bibitem[{{Helling} {et~al.}(2008{\natexlab{b}}){Helling}, {Dehn}, {Woitke}, \&
  {Hauschildt}}]{2008ApJ...675L.105H}
{Helling}, Ch., {Dehn}, M., {Woitke}, P., \& {Hauschildt}, P.~H.
  2008{\natexlab{b}}, \apjl, 675, L105

\bibitem[{{Helling} \& {Woitke}(2006)}]{2006A&A...455..325H}
{Helling}, Ch. \& {Woitke}, P. 2006, \aap, 455, 325

\bibitem[{{Helling} {et~al.}(2008{\natexlab{c}}){Helling}, {Woitke}, \&
  {Thi}}]{2008arXiv0803.4315H}
{Helling}, Ch., {Woitke}, P., \& {Thi}, W.~. 2008{\natexlab{c}}, ArXiv e-prints,
  803

\bibitem[{{Henize} \& {Mendoza}(1973)}]{1973ApJ...180..115H}
{Henize}, K.~G. \& {Mendoza}, E.~E. 1973, \apj, 180, 115

\bibitem[{{Joergens}(2006)}]{2006A&A...448..655J}
{Joergens}, V. 2006, \aap, 448, 655

\bibitem[{{Johnas} {et~al.}(2008){Johnas}, {Helling}, {Dehn}, {Woitke}, \&
  {Hauschildt}}]{2008MNRAS.385L.120J}
{Johnas}, C.~M.~S., {Helling}, Ch., {Dehn}, M., {Woitke}, P., \& {Hauschildt},
  P.~H. 2008, \mnras, 385, L120

\bibitem[{{Jung} {et~al.}(2006){Jung}, {Lundin}, {Modigliani}, {Dobrzycka}, \&
  {Hummel}}]{2006ASPC..351..295J}
{Jung}, Y., {Lundin}, L.~K., {Modigliani}, A., {Dobrzycka}, D., \& {Hummel}, W.
  2006, in Astronomical Society of the Pacific Conference Series, Vol. 351,
  Astronomical Data Analysis Software and Systems XV, ed. C.~{Gabriel},
  C.~{Arviset}, D.~{Ponz}, \& S.~{Enrique}, 295

\bibitem[{{Kainulainen} {et~al.}(2006){Kainulainen}, {Lehtinen}, \&
  {Harju}}]{2006A&A...447..597K}
{Kainulainen}, J., {Lehtinen}, K., \& {Harju}, J. 2006, \aap, 447, 597

\bibitem[{{Kenyon} \& {Hartmann}(1995)}]{1995ApJS..101..117K}
{Kenyon}, S.~J. \& {Hartmann}, L. 1995, \apjs, 101, 117

\bibitem[{{Kholopov} {et~al.}(1981){Kholopov}, {Samus'}, {Kukarkina},
  {Medvedeva}, \& {Perova}}]{1981IBVS.1921....1K}
{Kholopov}, P.~N., {Samus'}, N.~N., {Kukarkina}, N.~P., {Medvedeva}, G.~I., \&
  {Perova}, N.~B. 1981, Informational Bulletin on Variable Stars, 1921, 1

\bibitem[{{Kirkpatrick}(2005)}]{2005ARA&A..43..195K}
{Kirkpatrick}, J.~D. 2005, \araa, 43, 195

\bibitem[{{Kirkpatrick} {et~al.}(2006){Kirkpatrick}, {Barman}, {Burgasser},
  {McGovern}, {McLean}, {Tinney}, \& {Lowrance}}]{2006ApJ...639.1120K}
{Kirkpatrick}, J.~D., {Barman}, T.~S., {Burgasser}, A.~J., {et~al.} 2006, \apj,
  639, 1120

\bibitem[{{Kurosawa} {et~al.}(2005){Kurosawa}, {Harries}, \&
  {Symington}}]{2005MNRAS.358..671K}
{Kurosawa}, R., {Harries}, T.~J., \& {Symington}, N.~H. 2005, \mnras, 358, 671

\bibitem[{{Lafreniere} {et~al.}(2008){Lafreniere}, {Jayawardhana}, {Brandeker},
  {Ahmic}, \& {van Kerkwijk}}]{2008arXiv0803.0561L}
{Lafreniere}, D., {Jayawardhana}, R., {Brandeker}, A., {Ahmic}, M., \& {van
  Kerkwijk}, M.~H. 2008, ArXiv e-prints, 803

\bibitem[{{Lawrence} {et~al.}(2007){Lawrence}, {Warren}, {Almaini}, {Edge},
  {Hambly}, {Jameson}, {Lucas}, {Casali}, {Adamson}, {Dye}, {Emerson},
  {Foucaud}, {Hewett}, {Hirst}, {Hodgkin}, {Irwin}, {Lodieu}, {McMahon},
  {Simpson}, {Smail}, {Mortlock}, \& {Folger}}]{2007MNRAS.379.1599L}
{Lawrence}, A., {Warren}, S.~J., {Almaini}, O., {et~al.} 2007, \mnras, 379,
  1599

\bibitem[{{Lawson} {et~al.}(1996){Lawson}, {Feigelson}, \&
  {Huenemoerder}}]{1996MNRAS.280.1071L}
{Lawson}, W.~A., {Feigelson}, E.~D., \& {Huenemoerder}, D.~P. 1996, \mnras,
  280, 1071

\bibitem[{{Lenzen} {et~al.}(2003){Lenzen}, {Hartung}, {Brandner}, {Finger},
  {Hubin}, {Lacombe}, {Lagrange}, {Lehnert}, {Moorwood}, \&
  {Mouillet}}]{2003SPIE.4841..944L}
{Lenzen}, R., {Hartung}, M., {Brandner}, W., {et~al.} 2003, in Presented at the
  Society of Photo-Optical Instrumentation Engineers (SPIE) Conference, Vol.
  4841, Instrument Design and Performance for Optical/Infrared Ground-based
  Telescopes. Edited by Iye, Masanori; Moorwood, Alan F. M. Proceedings of the
  SPIE, Volume 4841, pp. 944-952 (2003)., ed. M.~{Iye} \& A.~F.~M. {Moorwood},
  944--952

\bibitem[{{Liu} {et~al.}(2003){Liu}, {Najita}, \&
  {Tokunaga}}]{2003ApJ...585..372L}
{Liu}, M.~C., {Najita}, J., \& {Tokunaga}, A.~T. 2003, \apj, 585, 372

\bibitem[{{Lodieu} {et~al.}(2008){Lodieu}, {Hambly}, {Jameson}, \&
  {Hodgkin}}]{2008MNRAS.383.1385L}
{Lodieu}, N., {Hambly}, N.~C., {Jameson}, R.~F., \& {Hodgkin}, S.~T. 2008,
  \mnras, 383, 1385

\bibitem[{{Luhman}(1999)}]{1999ApJ...525..466L}
{Luhman}, K.~L. 1999, \apj, 525, 466

\bibitem[{{Luhman}(2004)}]{2004ApJ...602..816L}
{Luhman}, K.~L. 2004, \apj, 602, 816

\bibitem[{{Luhman} \& {Muench}(2008)}]{2008arXiv0805.3722L}
{Luhman}, K.~L. \& {Muench}, A.~A. 2008, ArXiv e-prints, 805

\bibitem[{{Luhman} {et~al.}(2003){Luhman}, {Stauffer}, {Muench}, {Rieke},
  {Lada}, {Bouvier}, \& {Lada}}]{2003ApJ...593.1093L}
{Luhman}, K.~L., {Stauffer}, J.~R., {Muench}, A.~A., {et~al.} 2003, \apj, 593,
  1093

\bibitem[{{Luhman} {et~al.}(2006){Luhman}, {Wilson}, {Brandner}, {Skrutskie},
  {Nelson}, {Smith}, {Peterson}, {Cushing}, \& {Young}}]{2006ApJ...649..894L}
{Luhman}, K.~L., {Wilson}, J.~C., {Brandner}, W., {et~al.} 2006, \apj, 649, 894

\bibitem[{{Marley} {et~al.}(2007){Marley}, {Fortney}, {Hubickyj},
  {Bodenheimer}, \& {Lissauer}}]{2007ApJ...655..541M}
{Marley}, M.~S., {Fortney}, J.~J., {Hubickyj}, O., {Bodenheimer}, P., \&
  {Lissauer}, J.~J. 2007, \apj, 655, 541

\bibitem[{{Meyer} \& {Wilking}(2008)}]{2008Meyer}
{Meyer}, M.~R. \& {Wilking}, B.~A. 2008, in preparation

\bibitem[{{Mugrauer} \& {Neuh{\"a}user}(2005)}]{2005AN....326..701M}
{Mugrauer}, M. \& {Neuh{\"a}user}, R. 2005, Astronomische Nachrichten, 326, 701

\bibitem[{{Mugrauer} {et~al.}(2006){Mugrauer}, {Seifahrt}, {Neuh{\"a}user}, \&
  {Mazeh}}]{2006MNRAS.373L..31M}
{Mugrauer}, M., {Seifahrt}, A., {Neuh{\"a}user}, R., \& {Mazeh}, T. 2006,
  \mnras, 373, L31

\bibitem[{{Muzerolle} {et~al.}(1998){Muzerolle}, {Hartmann}, \&
  {Calvet}}]{1998AJ....116..455M}
{Muzerolle}, J., {Hartmann}, L., \& {Calvet}, N. 1998, \aj, 116, 455

\bibitem[{{Nagasawa} {et~al.}(2008){Nagasawa}, {Ida}, \&
  {Bessho}}]{2008ApJ...678..498N}
{Nagasawa}, M., {Ida}, S., \& {Bessho}, T. 2008, \apj, 678, 498

\bibitem[{{Nakajima} {et~al.}(1995){Nakajima}, {Oppenheimer}, {Kulkarni},
  {Golimowski}, {Matthews}, \& {Durrance}}]{1995Natur.378..463N}
{Nakajima}, T., {Oppenheimer}, B.~R., {Kulkarni}, S.~R., {et~al.} 1995, \nat,
  378, 463

\bibitem[{{Natta} {et~al.}(2000){Natta}, {Meyer}, \&
  {Beckwith}}]{2000ApJ...534..838N}
{Natta}, A., {Meyer}, M.~R., \& {Beckwith}, S.~V.~W. 2000, \apj, 534, 838

\bibitem[{{Natta} \& {Testi}(2001)}]{2001A&A...376L..22N}
{Natta}, A. \& {Testi}, L. 2001, \aap, 376, L22

\bibitem[{{Neuh{\"a}user} {et~al.}(2005){Neuh{\"a}user}, {Guenther},
  {Wuchterl}, {Mugrauer}, {Bedalov}, \& {Hauschildt}}]{2005A&A...435L..13N}
{Neuh{\"a}user}, R., {Guenther}, E.~W., {Wuchterl}, G., {et~al.} 2005, \aap,
  435, L13

\bibitem[{{Neuh{\"a}user} {et~al.}(2008){Neuh{\"a}user}, {Mugrauer},
  {Seifahrt}, {Schmidt}, \& {Vogt}}]{2008A&A...484..281N}
{Neuh{\"a}user}, R., {Mugrauer}, M., {Seifahrt}, A., {Schmidt}, T.~O.~B., \&
  {Vogt}, N. 2008, \aap, 484, 281

\bibitem[{{Oppenheimer} {et~al.}(1995){Oppenheimer}, {Kulkarni}, {Matthews}, \&
  {Nakajima}}]{1995Sci...270.1478O}
{Oppenheimer}, B.~R., {Kulkarni}, S.~R., {Matthews}, K., \& {Nakajima}, T.
  1995, Science, 270, 1478

\bibitem[{{Rieke} \& {Lebofsky}(1985)}]{1985ApJ...288..618R}
{Rieke}, G.~H. \& {Lebofsky}, M.~J. 1985, \apj, 288, 618

\bibitem[{{Rousset} {et~al.}(2003){Rousset}, {Lacombe}, {Puget}, {Hubin},
  {Gendron}, {Fusco}, {Arsenault}, {Charton}, {Feautrier}, {Gigan}, {Kern},
  {Lagrange}, {Madec}, {Mouillet}, {Rabaud}, {Rabou}, {Stadler}, \&
  {Zins}}]{2003SPIE.4839..140R}
{Rousset}, G., {Lacombe}, F., {Puget}, P., {et~al.} 2003, in Presented at the
  Society of Photo-Optical Instrumentation Engineers (SPIE) Conference, Vol.
  4839, Adaptive Optical System Technologies II. Edited by Wizinowich, Peter
  L.; Bonaccini, Domenico. Proceedings of the SPIE, Volume 4839, pp. 140-149
  (2003)., ed. P.~L. {Wizinowich} \& D.~{Bonaccini}, 140--149

\bibitem[{{Rydgren}(1980)}]{1980AJ.....85..444R}
{Rydgren}, A.~E. 1980, \aj, 85, 444

\bibitem[{{Santos} {et~al.}(2008){Santos}, {Melo}, {James}, {Gameiro},
  {Bouvier}, \& {Gomes}}]{2008A&A...480..889S}
{Santos}, N.~C., {Melo}, C., {James}, D.~J., {et~al.} 2008, \aap, 480, 889

\bibitem[{{Scholz} {et~al.}(2003){Scholz}, {McCaughrean}, {Lodieu}, \&
  {Kuhlbrodt}}]{2003A&A...398L..29S}
{Scholz}, R.-D., {McCaughrean}, M.~J., {Lodieu}, N., \& {Kuhlbrodt}, B. 2003,
  \aap, 398, L29

\bibitem[{{Seifahrt} {et~al.}(2007){Seifahrt}, {Neuh{\"a}user}, \&
  {Hauschildt}}]{2007A&A...463..309S}
{Seifahrt}, A., {Neuh{\"a}user}, R., \& {Hauschildt}, P.~H. 2007, \aap, 463,
  309

\bibitem[{{Slesnick} {et~al.}(2006){Slesnick}, {Carpenter}, \&
  {Hillenbrand}}]{2006AJ....131.3016S}
{Slesnick}, C.~L., {Carpenter}, J.~M., \& {Hillenbrand}, L.~A. 2006, \aj, 131,
  3016

\bibitem[{{Stassun} {et~al.}(2006){Stassun}, {Mathieu}, \&
  {Valenti}}]{2006Natur.440..311S}
{Stassun}, K.~G., {Mathieu}, R.~D., \& {Valenti}, J.~A. 2006, \nat, 440, 311

\bibitem[{{Stassun} {et~al.}(2007){Stassun}, {Mathieu}, \&
  {Valenti}}]{2007ApJ...664.1154S}
{Stassun}, K.~G., {Mathieu}, R.~D., \& {Valenti}, J.~A. 2007, \apj, 664, 1154

\bibitem[{{Tsuji}(2005)}]{2005ApJ...621.1033T}
{Tsuji}, T. 2005, \apj, 621, 1033

\bibitem[{{Tsuji} {et~al.}(1996){Tsuji}, {Ohnaka}, {Aoki}, \&
  {Nakajima}}]{1996A&A...308L..29T}
{Tsuji}, T., {Ohnaka}, K., {Aoki}, W., \& {Nakajima}, T. 1996, \aap, 308, L29

\bibitem[{{Weinberg} {et~al.}(1987){Weinberg}, {Shapiro}, \&
  {Wasserman}}]{1987ApJ...312..367W}
{Weinberg}, M.~D., {Shapiro}, S.~L., \& {Wasserman}, I. 1987, \apj, 312, 367

\bibitem[{{Weintraub}(1990)}]{1990ApJS...74..575W}
{Weintraub}, D.~A. 1990, \apjs, 74, 575

\bibitem[{{Whelan} {et~al.}(2004){Whelan}, {Ray}, \&
  {Davis}}]{2004A&A...417..247W}
{Whelan}, E.~T., {Ray}, T.~P., \& {Davis}, C.~J. 2004, \aap, 417, 247

\bibitem[{{Whittet} {et~al.}(1987){Whittet}, {Kirrane}, {Kilkenny}, {Oates},
  {Watson}, \& {King}}]{1987MNRAS.224..497W}
{Whittet}, D.~C.~B., {Kirrane}, T.~M., {Kilkenny}, D., {et~al.} 1987, \mnras,
  224, 497

\bibitem[{{Whittet} {et~al.}(1997){Whittet}, {Prusti}, {Franco}, {Gerakines},
  {Kilkenny}, {Larson}, \& {Wesselius}}]{1997A&A...327.1194W}
{Whittet}, D.~C.~B., {Prusti}, T., {Franco}, G.~A.~P., {et~al.} 1997, \aap,
  327, 1194

\bibitem[{{Wilson} {et~al.}(2003){Wilson}, {Miller}, {Gizis}, {Skrutskie},
  {Houck}, {Kirkpatrick}, {Burgasser}, \& {Monet}}]{2003IAUS..211..197W}
{Wilson}, J.~C., {Miller}, N.~A., {Gizis}, J.~E., {et~al.} 2003, in IAU
  Symposium, Vol. 211, Brown Dwarfs, ed. E.~{Mart{\'{\i}}n}, 197--+

\bibitem[{{Woitke} \& {Helling}(2003)}]{2003A&A...399..297W}
{Woitke}, P. \& {Helling}, Ch. 2003, \aap, 399, 297

\bibitem[{{Woitke} \& {Helling}(2004)}]{2004A&A...414..335W}
{Woitke}, P. \& {Helling}, Ch. 2004, \aap, 414, 335

\bibitem[{{Zacharias} {et~al.}(2004){Zacharias}, {Urban}, {Zacharias},
  {Wycoff}, {Hall}, {Monet}, \& {Rafferty}}]{2004AJ....127.3043Z}
{Zacharias}, N., {Urban}, S.~E., {Zacharias}, M.~I., {et~al.} 2004, \aj, 127,
  3043

\end{thebibliography}

\onlfig{4}{
\begin{figure*}
   \resizebox{\hsize}{!}{\includegraphics{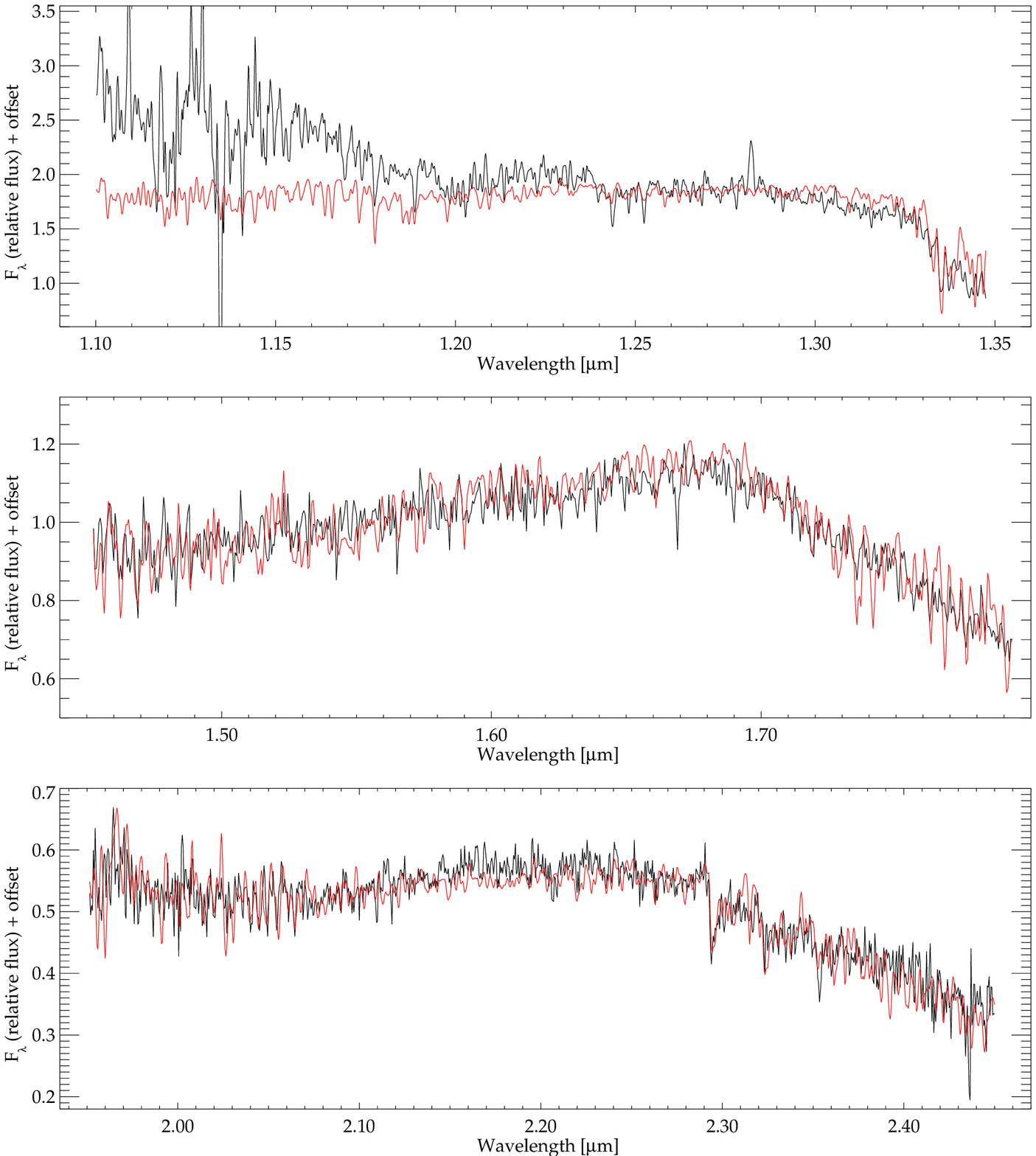}}
   \caption{\textit{From top to bottom:} J, H, and K band spectra, adjusted in their relative flux levels. \textit{In each panel:} Our SINFONI spectra of the CT Cha companion in spectral resolution 1500 are shown (\textit{black}) in comparison to the best-fitting Drift-Phoenix \citep{2008ApJ...675L.105H} synthetic spectrum (\textit{red}, same spectral resolution) of 2600 K, $\log{g}$=\,3.5, [Me/H]=\,0.0, and a visual extinction of A$_{V}$=\,5.2 mag. Note the strong Pa-$\beta$ emission line at $\sim$\,1.282 $\mu$m in the J-band and the weak alkali lines in the J-band (\ion{K}{I},$\sim$\,1.25 $\mu$m and \ion{Al}{I}, $\sim$\,1.315 $\mu$m) and the K-band (Na I doublet, $\sim$\,1.25 $\mu$m) of the synthetic model. This, as well as the overshoot of the water vapor bands, e.g. in the blue part of the J-band, are currently in the process of investigation \citep[see e.g.][]{2008MNRAS.385L.120J}. Several signs of youth are present in the spectrum of the CT Cha companion, like e.g. the perfectly triangular H-band. See text for more details.}
   \label{spectraonline_CTCha}
   \end{figure*}
}

\onlfig{6}{
\begin{figure*}
   \resizebox{\hsize}{!}{\includegraphics{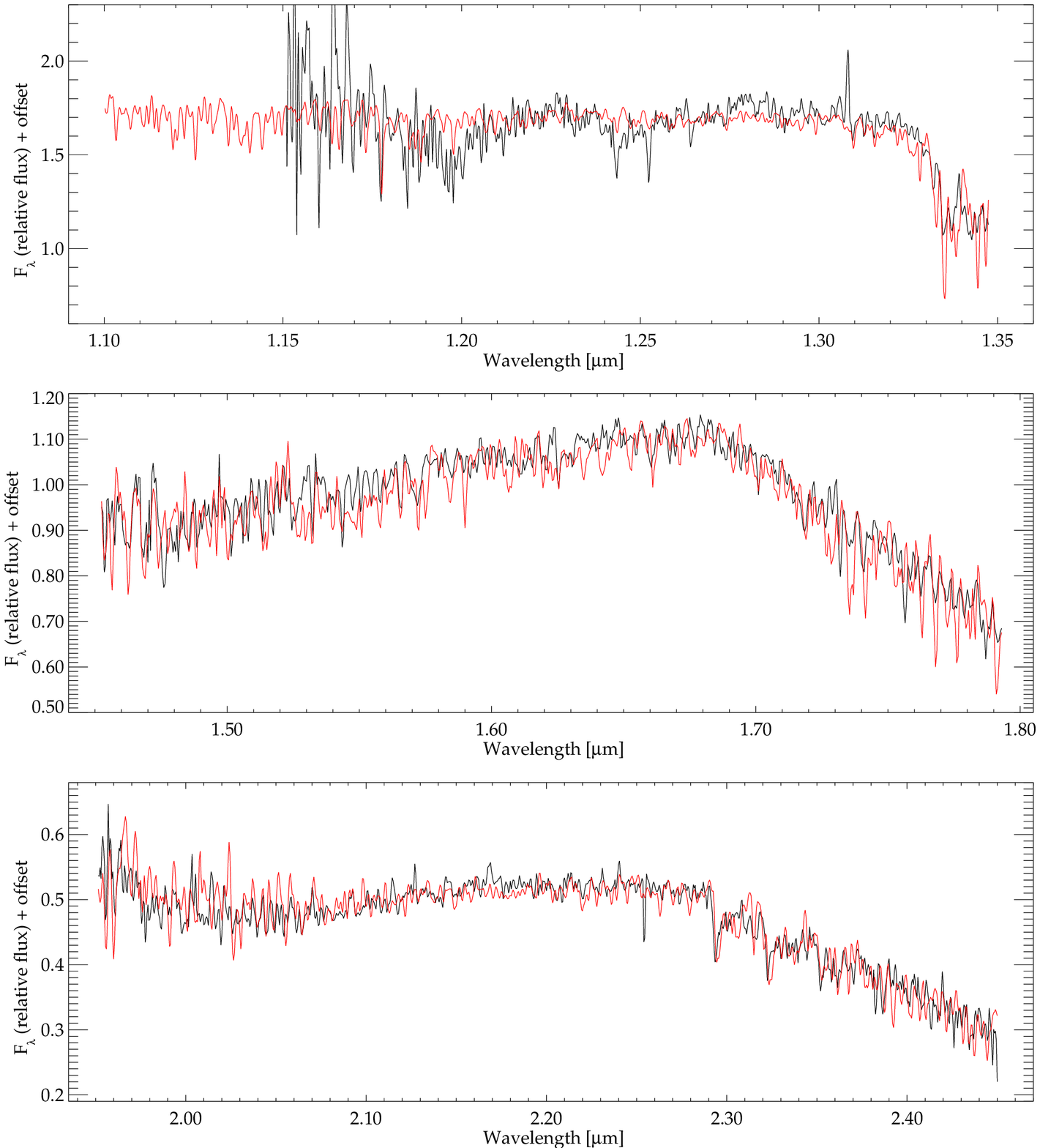}}
   \caption{\textit{From top to bottom:} J, H, and K band spectra, adjusted in their relative flux levels. \textit{In each panel:} The spectrum of the free-floating brown dwarf USco J160648-223040 \citet{2008MNRAS.383.1385L} in spectral resolution 1500 is shown (\textit{black}) in comparison to the best-fitting Drift-Phoenix \citep{2008ApJ...675L.105H} synthetic spectrum (\textit{red}, same spectral resolution) of 2700 K, $\log{g}$=\,3.5, [Me/H]=\,-0.5 and a visual extinction of A$_{V}$=\,0.2 mag. Shown is the full available spectrum of USco J160648-223040 beginning at 1.15\,$\mu$m. There are some peaked deviations in the object's spectrum in contrast to the model, probably caused by detected cosmic rays. See text for more details.}
   \label{spectraonline_UScoJ160648-223040}
   \end{figure*}
}

\onlfig{8}{
\begin{figure*}
   \resizebox{\hsize}{!}{\includegraphics{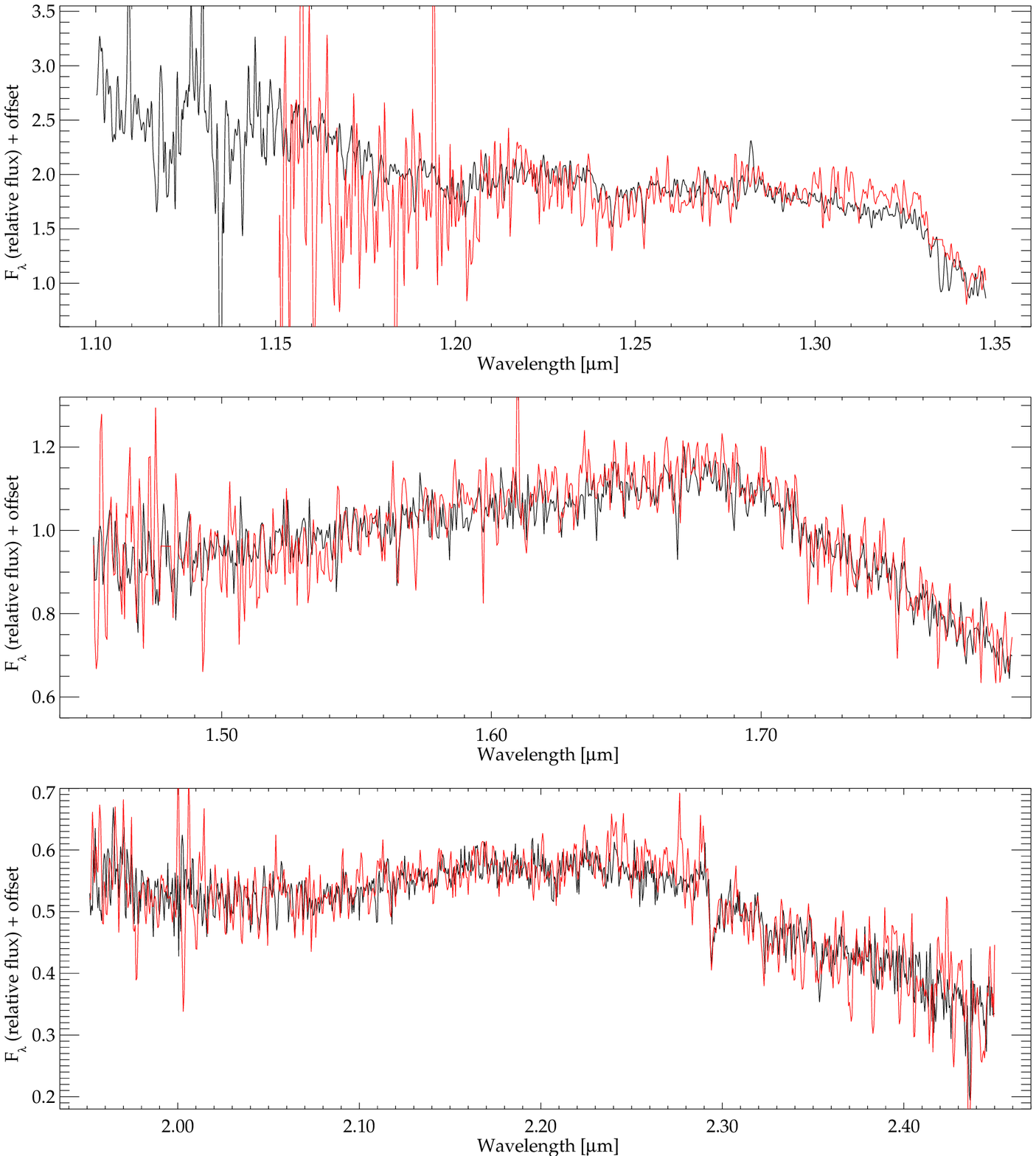}}
   \caption{\textit{From top to bottom:} J, H, and K band spectra, adjusted in their relative flux levels. \textit{In each panel:} Our SINFONI spectra of the CT Cha companion in spectral resolution 1500 (\textit{black}) are shown in comparison to the spectrum of the free-floating brown dwarf USco J160714-232101 \citep{2008MNRAS.383.1385L} (\textit{red}, same spectral resolution). The fit of both objects with Drift-Phoenix \citep{2008ApJ...675L.105H} gives 2600 K, $\log{g}$=\,3.5, [Me/H]=\,0.0, while they have A$_{V}$=\,5.2\,mag and A$_{V}$=\,2.1\,mag respectively. Shown is the full available spectrum of USco J160714-232101 beginning at 1.15\,$\mu$m. There are some peaked deviations in USco J160714-232101's spectrum in contrast to the CT Cha companion candidate, probably caused by detected cosmic rays. Further note the difference in the Pa-$\beta$ emission line at $\sim$\,1.282 $\mu$m. See text for more details.}
   \label{spectraonline_UScoJ160714-232101}
   \end{figure*}
}

\onlfig{11}{
\begin{figure*}
   \resizebox{\hsize}{!}{\includegraphics{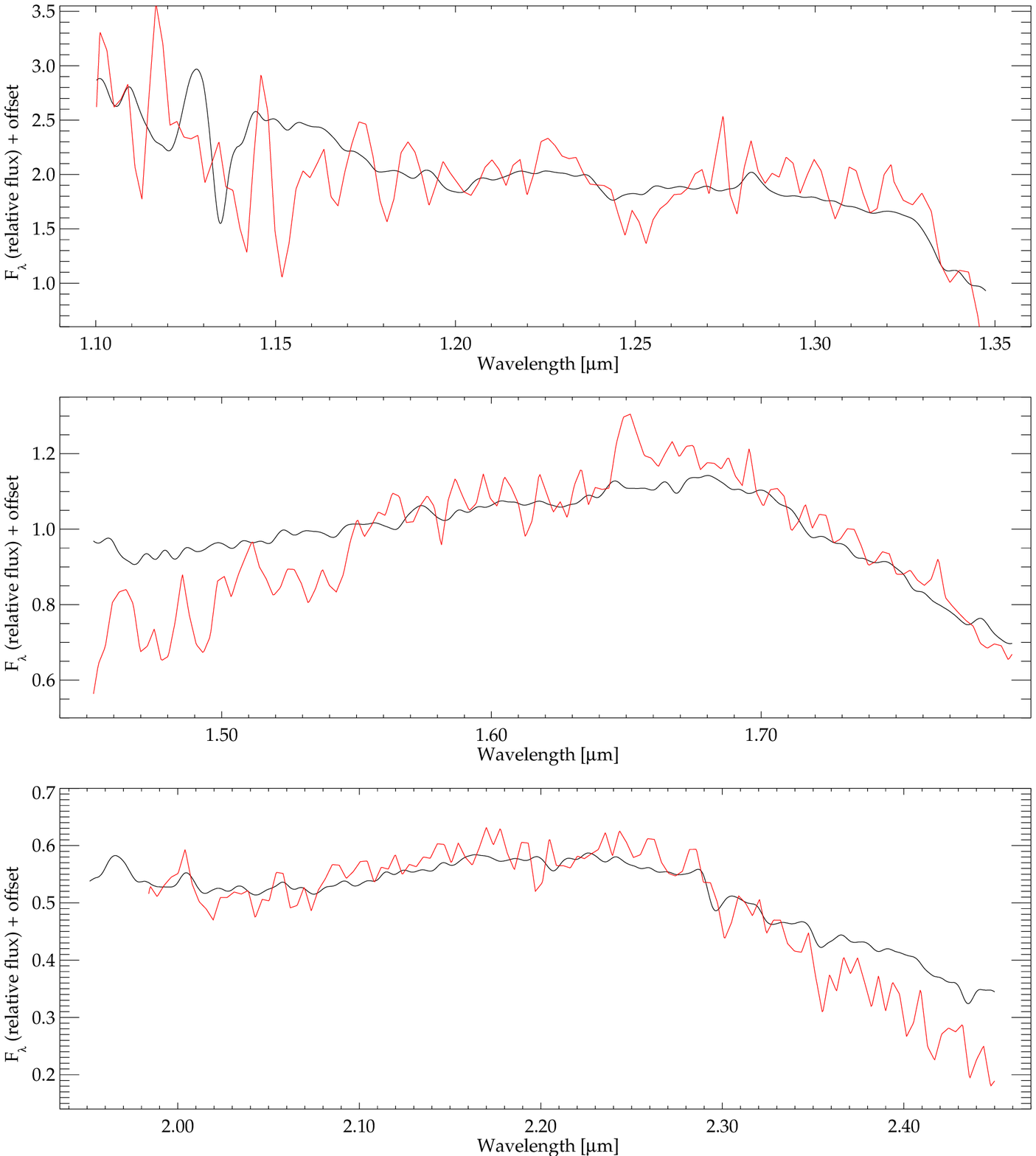}}
   \caption{\textit{From top to bottom:} J, H, and K band spectra, adjusted in their relative flux levels. \textit{In each panel:} Our SINFONI spectra of the CT Cha companion in spectral resolution 300 (\textit{black}) are shown in comparison to the spectrum of the companion CHXR 73 B  \citet{2006ApJ...649..894L}, also a member of Cha I (\textit{red}, same spectral resolution). The fit of both objects with Drift-Phoenix \citep{2008ApJ...675L.105H} gives 2600 K, [Me/H]=\,0.0, while they have A$_{V}$=\,5.2\,mag,  $\log{g}$=\,3.5, and A$_{V}$=\,12.6\,mag, $\log{g}$=\,4.5 respectively. Note the deviant slopes in the blue part of the H-band and the red part of the K-band, possibly caused by the lower signal-to-noise in these parts of the bands, as can be seen in Fig.~\ref{Dered_CHXR73B}. See text for more details.}
   \label{spectraonline_CHXR73B}
   \end{figure*}
}

\end{document}